\documentclass[11pt]{article} 

\usepackage{amsmath,amsthm,latexsym,amssymb,amsfonts,epsfig}

\newtheorem*{Theorem}{Theorem}

\newcommand{\be}{\begin{equation}}
\newcommand{\ee}{\end{equation}}
\newcommand{\ba}{\begin{eqnarray}}
\newcommand{\ea}{\end{eqnarray}}

\title{{\sf Quantum Cosmological Backreactions I:}\\
{\sf Cosmological Space Adiabatic Perturbation Theory}} 
\author{
{\sf S. Schander}$^1$\thanks{{\sf 
susanne.schander@gravity.fau.de}},
{\sf T. Thiemann}$^1$\thanks{{\sf 
thomas.thiemann@gravity.fau.de}}\\
\\
{\sf $^1$ Inst. for Quantum Gravity, FAU Erlangen -- N\"urnberg,}\\
{\sf Staudtstr. 7, 91058 Erlangen, Germany}\\
}
\date{{\small\sf \today}}

\makeatletter
\@addtoreset{equation}{section}
\makeatother

\begin{document} 

\maketitle

{\sf

\begin{abstract}
The search for quantum gravity 
fingerprints in currently available cosmological data that have 
their origin from the Planck era is of growing interest due to major recent
progress both in the theoretical modelling as well as the observational 
precision. Unsurprisingly, the theoretical predictions are very sensitive 
to the quantum effects that occur close to the classical big bang 
singularity. It is therefore of substantial interest to describe these 
effects as precisely as possible.

This is the first in a series of papers that aim at 
improving  on the treatment of quantum 
effects that arise due to backreactions between matter and geometry. 
The technique we employ is space adiabatic perturbation theory (SAPT)
in the form developed in seminal papers by Panati, Spohn and Teufel.
SAPT is a generalisation of the more familiar Born Oppenheimer Approximation
(BOA) that applies well in systems that allow a split of the degrees of 
freedom into two sets that propagate on rather different time scales
such as the homogeneous and inhomogeneous field modes in cosmology.
We will show that this leads to presently neglected correction terms 
in the quantum Friedman equations.   

In the present paper we adapt and generalise SAPT to the hybrid approach to 
quantum cosmology developed by Mena Marugan et al. that allows for a 
systematic quantum separation of the (in)homogeneous modes.  
Since SAPT was developed for
quantum mechanics rather than quantum field theory, several challenges have 
to be met.  
\end{abstract}

\newpage

\tableofcontents

\newpage

\section{Introduction}
\label{s1}

Physical systems often display a separation of time scales which can be 
exploited when solving the equations governing them. The historically 
first example was the approximate solution of the spectral problem for 
molecules introduced by Born and Oppenheimer \cite{1} henceforth 
called Born Oppenheimer approximation (BOA). Here the slow
degrees of freedom are the nuclear ones while the fast degrees of freedom 
are those of the electrons owing to a mass ratio of about 2000 or more. 
Intuitively then the electrons "adiabatically" adapt their fast motion
to the slow motion of the nuclear degrees of freedom without much disturbance.

In the leading order of that mass ratio, one neglects the modifications that 
the nuclear part of the Hamiltonian has on the eigenstates of the electrons 
which parametrically depend on the location of the nuclear degress of freedom.
This allows for an approximate solution of the full spectral problem in that 
one finds an effective Hamiltonian for the nuclear degrees of freedom 
on a subspace of the full Hilbert space labelled 
by an eigenvalue of the electronic part of the Hamiltonian leading 
to the {\it energy bands} of the coupled system.
However, when one tries to go beyond that leading order and tries to 
incorporate {\it backreaction effects}, one is confronted with an 
infinite set of coupled equations labelled by the undisturbed 
energy eigenvalues 
of the electrons which in contrast to the leading order is not solvable 
exactly in practice. 

One way to view space adiabatic perturbation theory (SAPT), developed 
in particular by Panati, Spohn and Teufel (see \cite{2} and references 
therein), is to consider 
it to be as a systematic approach to perturbatively decouple this infinite
set of equations. Namely, the backreaction mixes the undisturbed electronic 
energy eigenstates and the exact effective Hamiltonian acts on a subspace
of the Hilbert space which is rotated as compared to the undisturbed subspace
labelled by the energy band parameter. SAPT provides a systematic 
perturbative construction, for each energy level of the unperturbed 
fast part of the Hamiltonian, of i. the projection onto the corresponding 
subspace, ii. the unitary rotation between that subspace and the undisturbed
one and iii. the effective slow Hamiltonian acting on that subspace.
Being a perturbation series, the corresponding operators can be truncated 
at the desired level of precision.       

This ``space'' adiabatic scheme, that we just described, includes the  
``time'' adiabatic one (see e.g. \cite{3}) which was developed for explicitly 
time dependent 
Hamiltonians, where the external time scale is much larger than the 
internal one, in the following sense: We can pass to the extended phase space 
formalism and treat time as a dynamical ``spatial'' degree of freedom 
if we impose the constraint that the sum of Hamiltonian and momentum conjugate 
to time is vanishing. As such, SAPT is already well adapted to generally
covariant systems that have no Hamiltonian but rather a Hamiltonian 
constraint as it happens in quantum gravity and more specifically 
quantum comsmology. 

In this series of papers we thus advocate to further develop 
the work of \cite{14a} and employ SAPT for 
quantum cosmology, in particular quantum cosmological perturbation theory
(QCPT). To make the presentation as clear as possible, we consider 
as matter content just an inflaton field (and possibly Gaussian dust 
as a dynamical clock, see \cite{4,5} and references therein) but 
our formalism can be easily extended to more complicated models.
To see how SAPT can be applied, note that in cosmological 
perturbation theory one splits the field degrees of freedom into a 
dominant, homogeneous contribution and an inhomogeneous perturbation 
thereof \cite{6}. The corresponding perturbation series in principle 
involves all orders, but already the second order is non trivial and 
we confine ourselves to second order in this series of papers.
In particular, it makes a crucial difference whether one considers 
the homogeneous sector as a background or not. In either case one must 
deal with the issue of gauge invariance with respect to the perturbative
fragment of the spacetime diffeomorphism group. In the background approach,
which by definition neglects backreaction effects, one considers only the 
inhomogeneous degrees of freedom as dynamical and in the quantum theory
thus effectively conducts quantum field theory on the given background 
spacetime \cite{7} (QFT in CST), albeit in terms of the corresponding 
gauge invariant 
field modes (e.g. Mukhanov-Sasaki field \cite{6} if there is no dust 
present). 

If one wants to treat the homogeneous degrees of freedom  
as dynamical and thus allows for backreaction effects then in particular
the constraint analysis must be properly adapted to extract the correct 
gauge invariant degrees of freedom. To the best of our knowledge, 
this clasiscal programme has been carried out for the first time (to second 
order) in \cite{8} and we will adopt it for our purposes. In particular 
it is important to carry out a canonical transformation (to second order) 
on the {\it full phase space} in order that the constraints still
form a first class system (to second order). Simultaneously that 
transformation must be chosen in such a way, that the second order 
contribution to the Hamiltonian (constraint) can be represented on an 
appropriate Hilbert space. 

When quantising the complete system, i.e. homogeneous and inhomogeneous 
degrees of freedom, one has the freedom to choose different types of 
Hilbert space representations for these two sets of variables. This 
is the idea of the {\it hybrid LQC} quantisation \cite{8}: Here 
the perturbations are represented on suitable Fock spaces, thus 
taking advantage of their at most quadratic appearance in the Hamiltonian 
constraint, while the homogeneous degrees of freedom are quantised by 
using Loop Qauntum Cosmology (LQC) \cite{9} techniques. Here LQC
is a mini supersapce quantisation of just the homogeneous degrees of freedom 
using techniques from Loop Quantum Gravity (LQG) \cite{10} which is 
a candidate theory for full quantum gravity. The LQC representation is well
adapted to the non-polymnomial appearance of the homogeneous degrees of 
freedom in the Hamiltonian (constraint). In this series of papers 
we will adopt parts of the hybrid LQC idea but we will consider different 
possibilities concerning the quantisation of the homogeneous sector.

The current ideas, mostly within the LQC scenario, 
to describe the interaction between homogeneous and inhomogeneoeus 
degrees of freedom include:
1. the dressed metric approach \cite{11}, 2. the rainbow metric approach 
\cite{12}, the 3. the deformed algebra 
approach \cite{13}\ and the hybrid approach \cite{8}.
In  the dressed metric approach, for full LQG proposed for the first time 
in \cite{14},
one computes the partial expectation values 
of the Hamiltonian (constraint) with respect to a semiclassical state 
of the homogeneous sector and then derives the free QFT in CST on the resulting
background spacetime. The dynamics of the background is chosen to be 
derived from the expectation value of the homogeneous contribution 
to the Hamiltonian constraint with respect to that semiclassical state 
(including quantum corrections from fluctuations and those that come 
from the particular LQC quantisation method), 
leading to the so called {\it effective equations}. In the rainbow metric 
approach one similarly takes such a partial expectation value and then recasts
the resulting effective metric, separately for each term of the Hamiltonian 
(constraint) labelled by a mode number, into a FRW form thereby making 
scale factor and lapse function depending on that mode number (hence the name 
rainbow). As far as the quantisation  of the inhomogeneous sector is 
concrened, both of these approaches are equivalent. In the deformed 
algebra approach one requires that the constraints up to second order in the 
inhomogneities close (up to second order) in the sense of Poisson brackets 
when one replaces the background 
variables by effective functions thereof. This can be considered as 
a different method to choose the semiclassical state although it is not clear 
that a semiclassical state exists which reproduces all the coefficients 
neeeded for such a closure. Finally, in the hybrid approach, one assumes 
that certain quantum states for the homogeneous sector exist such that the 
full quantum constraint including second order inhomogenous contributions 
can be recast into a Schr\"odinger type first order equation with respect to the 
homogenous inflaton mode considered as an internal time. The list of assumptions 
include that second order derivatives with respect to internal
time can be neglected and the validity of an Ansatz for the wave functions 
of Born-Oppenheimer type.   

To the best of our understanding, these methods do not 
incorporate backreaction effects 
in the same sense as the zeroth order of the BOA does not include them. Also,
to the best of our understanding,
the various assumptions that went into these approximations are not easy
to control.  
Accordingly, it is an important question how one can improve on this.
In this series of papers we would like to convince the reader that 
SAPT methods are a powerful tool for achieving just that. SAPT follows an 
iterative systematic algorithm that can be applied as easily as standard quantum 
mechanical stationary perturbation theory can be, although concrete 
calculations become quickly involved and tedious as we increase the adiabatic 
order. The basic idea is that 
the homogeneous modes can be considered as the slow degrees of freedom 
while the inhomogeneous ones play the role of the fast degrees of 
freedom. Roughly this happens because the homogeneous mode, being 
the integral (or sum) over the inhomogeneous modes, is a kind 
of centre of mass mode with respect to the inhomogeneous ones and 
as in classical mechanics comes with the large total mass of the system 
rather than the small individual ones. This will be explained in later 
parts of this paper. Accordingly the SAPT scheme should be quite appropriate
for the hybrid treatment suggested in \cite{8}. 

However, the application of the SAPT scheme to quantum cosmology is not 
entirely straightforward. We meet the following challenges:\\
i.\\ 
SAPT was developed for systems with a finite number of degrees of freedom
while we are interested in quantum field theory. Hence, rather innocent
looking assumptions of the SAPT scheme such as that the Hilbert space 
can be considered as a tensor product of a fast and slow Hilbert space 
are no longer granted to make any sense as was first remarked in 
\cite{14a}.\\
ii.\\
Furthermore, many of the 
theorems proved in \cite{2} rely on the assumption of dealing with
everywhere smooth symbol classes of fast Hilbert space operator valued 
functions on the slow phase space which is not the case in our applications. Accordinly 
we will not have much to say about the convergence of the perturbation
series (in terms of the adiabatic parameter) and we leave that part
of the analysis for future research.\\
iii.\\
In contrast to BOA, in SAPT one is able to to deal with situations where 
the part of the Hamiltonian that describes the fast degrees of freedom 
depends on both configuration and momentum variables of the slow sector.
This is important for quantum cosmology since for instance the 
Mukhanov-Sasaki mass term has precisely this property and thus requires 
the full {\it Weyl quantisation technology} of the SAPT scheme. \\
iv.\\
Unfortunately, the mass squared terms that appear in quantum cosmology 
not only depend on negative powers of the configuration degrees of freedom 
of the homogeneous sector but also on the momenta. Even worse, they are 
not positive definite. This raises complicated domain issues 
both in the inhomogeneous QFT sector as well as in the homogeneous QM 
sector. In the QFT sector it begs the question of how to treat 
quantum fields with varying 
indefinite mass squared 
terms and in the QM sector one meets rather non-polynmial operators usually not 
discussed in QM and whose degree of non-polynomiality exceeds even the 
situation one meets in LQC.\\ 
\\
We also should comment on the physical intution of the adiabatic scheme, 
which is different for systems with Hamiltonian constraints and a true 
Hamiltionian respectively. In the more familiar latter case, the 
physical argument is based on 1. the equipartition theorem \cite{16}, 
2. the 
ergodicity assumption and 3. the assumption that the Hamiltonian
is of second order in the momenta. In that case the time average, which 
due to ergodicity equals the phase space average in say the canonical 
ensemble, of the kinetic terms in the Hamiltonian are all equal implying that
the light degrees of freedom in average are much faster than the heavy ones.
That intuition fails when we have a Hamiltonian constraint. However, in 
this case the constraint itself dictatates that the kinetic term 
of the slow mode is equal to the sum of the kinetic terms of the fast modes
which at least in the kinetic energy dominated regime of tyhe phase space 
leads to the same conclusion if the number of fast modes is much smaller 
than the inverse square of the adiabatic parameter. In a quantum mechanical
system that latter number is finite by itself in absolute terms, in the 
quantum field theory case it is effectively finite in any Fock state which 
only contains finitely many excitations and the reasoning will apply at least
to the lower lying excited states. 

On the other hand, technically speaking, the SAPT scheme works as soon as 
the system involves an adiabatic parameter that multiplies the 
momenta of the ``slow'' degrees of frteedom.\\
\\
The architecture of this paper is as follows:\\
\\
In section two we will derive the SAPT programme in a self-contained 
fashion for the case of a finite dimensional phase space. This also serves 
to introduce a simple notation aiming at highest possible transpararency.
     
In section three we will address the conceptual and mathematical 
complications when applying SAPT to quantum cosmology and their 
possible solutions, thus preparing the 
ground for the remaining papers of this series. 

In section four we summarise and give an outlook to the applications 
discussed in the other papers of this series.

\section{Elements of Space Adiabatic Perturbation Theory (SAPT)}
\label{s2}

In the first subsection we explain our notation and the basic 
ideas underlying the SAPT scheme. 
That notation is simplified as compared to the more technical subsequent papers of this 
series in order to be able to better focus on the underlying ideas. 
In the second subsection
we derive the essential inductive formulae underlying the 
adiabatic expansion. As mentioned before, we will not discuss the sense of 
convergence of the adiabatic perturbation series. This can be 
done by introducing notions from the theory of pseudo differential operators 
called symbol classes for which we refer 
the interested reader to \cite{2}.

\subsection{Notation and basic idea}
\label{s2.1}

Throughout 
this section we only consider finite dimensional phase spaces. To 
avoid unnecessary cluttering of formulae, 
it will in fact be sufficient to consider a four dimensional phase space, 
the generalisation 
to higher dimensional ones being straightworward and obvious. To a certain 
extent it is also possible to generalise this to finite dimensional 
phase spaces which are not vector spces \cite{14a}. 
The present phase space is thus  
coordinatised by a ``slow'' canonical pair $z=(q,p)$ and a fast canonical 
pair $(x,y)$ with standard canonical brackets
$\{p,q\}=\{y,x\}=1$, all others vanishing. The dimensionfree 
adiabatic parameter will be denoted by $\epsilon$. We denote the trivially 
rescaled slow momentum by $p':=\epsilon p$ and accordingly
$z'=(q,p')$. A basic assumption is 
that the Hamiltonian (constraint) of the system can be written in the form
\be \label{2.1.0}
h(q,p,x,y)=h_0((q,p',x,y;\epsilon)+\epsilon^r h_1(q,p',x,y\;\epsilon)
\ee
The term $h_1$ is allowed to vanish while $h_0$ never vanishes. If $h_1$ 
is non-vanishing we have $r\in \mathbb{N}-\{0\}$ and both 
$h_0, h_1$ are polynomials 
in $\epsilon$ of finite degree with coefficients independent of $\epsilon$
where the coefficients of zeroth order are non-vanishing. The form 
(\ref{2.1.0}) can often be obtained by multiplying the original Hamiltonian 
by a sufficiently high power of $\epsilon$ in which case the spectrum
of (\ref{2.1.0}) has to be rescaled by the corresponding inverse power, 
after the perturative computation has been completed. Furtheremore, the 
piece $h_0$ of the Hamiltonian comprises all terms of $h$ which 
upon quantisation of just the fast degrees of freedom allows for an 
easy diagonalisation on the Hilbert space 
${\cal H}_f$ of the fast degrees of freedom 
for all parameter values of $q,p'$.
Note that this means that any term of the form $\epsilon^s f(q,p'),\; n\ge 0$
in the classical Hamiltonian which is independent of $x,y$ will be subsumed 
under $h_0$ since the quantisation on ${\cal H}_f$ yields the 
trivial symbol $\epsilon^s f(q,p') 1_s$ and shifts the spectrum of 
$h_0(q,p')$ by $\epsilon^n f(q,p')$. This is in contrast to the BOA 
which would treat such a term as of higher order if either $n>0$ or 
$f(q,p')=\epsilon^l f(q,p),\; l>0$.   

Turning to the quantisation, we represent the ``slow'' degrees of freedom as 
operators $Q,P'=\epsilon P$ on the Hilbert space 
${\cal H}_s=L_2(\mathbb{R},dq)$ 
and the fast degrees of freedom as operators $X,Y$ on the Hilbert space 
${\cal H}_f=L_2(\mathbb{R},dx)$. The full Hilbert space is the 
tensor product 
Hilbert space ${\cal H}_f\otimes {\cal H}_s$ on which $x$ acts 
as $X\otimes 1_s$ and $q$ as $1_f\otimes Q$ respectively etc. 

We will also be dealing with so-called {\it symbols}.  These are functions 
on the slow phase space with values in the set of linear operators on 
${\cal H}_f$. We will assume that all symbols that we encounter 
have a common, invariant and dense domain for all $q,p'$ and are 
smooth in $q,p'$. This turns out to be the case in our applications 
as the symbols that we encounter are either those corresponding to 
(\ref{2.1.0}) or originate from the eigenfunctions $e_{n,a}(z')$ 
discussed below which both have the required properties.

Intuitively, a symbol $(q,p')\mapsto f(q,p')$ can be considered as 
arising from a function $f(q,p',x,y)$ on the full phase space by just 
quantising the fast sector and choosing some operator ordering 
to obtain $f(q,p')=f(q,p',X,Y)$. Unless confusion may arise, we will 
denote symbols and their underlying phase space functions by the same 
{\it lower case letter} $f$.

The next ingredient will be the {\it Weyl quantisation} $F:=W(f)$ of a 
symbol. We will denote Weyl quantisations by {\it capital letters} $F$.
These are now operators on the total product Hilbert space and in case 
that $f$ deopends polynomially on $p'$ then $F$ is just the symmetric 
ordering of the formal expression $f(Q,P')$. To be clear, the Weyl 
quantisation and the associated {\it Moyal product} is here with respect to 
the {\it rescaled Planck constant} $\hbar':=\epsilon \hbar$. This arises 
because of the commutation relations $[P',Q]=i\hbar' 1_s$ naturally as 
follows:\\
The Weyl elements are defined by
\be \label{2.1.1}
W(k,l):=\exp(i\frac{k\;Q+l P'}{\hbar'})
\ee
and the Weyl quantisation of a symbol $f$ by
\be \label{2.1.2}
W(f):=\int_{\mathbb{R}^2}\; \frac{dk\; dl}{[2\pi\hbar']^2}\;
\hat{f}(k,l)\otimes W(k,l)
\ee
where $\hat{f}$ denotes the Fourier transform of the symbol $f$
\be \label{2.1.3}
\hat{f}(k,l)=\int_{\mathbb{R}^2}\; dq\; dp'\; \exp(-i\frac{kq+lp'}{\hbar'})
\; f(q,p')
\ee
Note that $k,l$ have the dual dimension of $q,p'$ so that the products 
$kq,lp'$ have the dimension of $\hbar$. Then it is well known that for 
two symbols $f,g$ we have the {\it Moyal product} formula 
\ba \label{2.1.4}
W(f)\; W(g) &=&W(f\ast g),\; (f\ast g)(q,p'):=[\exp(\frac{i\hbar'}{2}
\theta_{12})\cdot f(q_1,p_1')\; g(q_2,p_2')]_{q_1=q_2=q,p_1'=p_2'=p'}
\nonumber\\
\theta_{12} &:=&
\frac{\partial^2}{\partial p_1' \partial q_2}      
-\frac{\partial^2}{\partial p_2' \partial q_1}      
\ea
The Moyal product is associative but not commutative. Note that in the 
{\it Moyal commutator}
\be \label{2.1.5}
[f,g]_\ast=f\ast g -g_\ast f=[f,g]+O(\hbar') 
\ee 
the  term $[f,g]= f\; g-g\; f$ of zeroth order in $\hbar'$ which is just the 
usual commutator of symbols is {\it not} vanishing in general in contrast
to the classical case of Weyl quantisation on just ${\cal H}_s$. 

Weyl quantisation thus serves several purposes at the same time:
First it allows us to write any operator on the full Hilbert space
${\cal H}_f\otimes {\cal H}_s$ in the form of a slow phase space  integral 
of elementary 
operators of the form $A_f(z)\otimes B_s(z)$ thus allowing us to simplify 
its spectral problem when the spectral problem of the $A_f(z)$ is known 
in closed form. Secondly, the Moyal product allows for a systematic 
power expansion in terms of the adiabatic parameter $\epsilon$ and 
thus enables us to set up a perturbative diagonalisation scheme. This
follows from the fact that $W(f) W(g)-W(fg)=W(f\ast g-fg)=O(\epsilon)$. 

Thus in what follows we will assume that the spectral problem of the 
symbols 
\be \label{2.1.6}
h_0(z):=h_0(z,X,Y;\epsilon)
\ee
is known in closed form. In our applications the spectra
will turn out to be pure 
point with eigenvalues $E_n(z'),\; n\in \mathbb{N}$ and an
orthonormal basis of eigenfunctions 
$e_{n,a}(z'),\; a=1,..,d_n$ where $d_n\in\mathbb{N}$ denotes its degeneracy
which we assume to be a constant in $z'$. 
An important assumption is the {\it absence of eigenvalue crossing},
that is, $E_m(z')-E_n(z')\not=0$ for all $z',\;m\not=n$. To get strong
mathematical results one may impose the stronger {\it gap condition} 
$g_n:=\inf_{z,m\not=n} |E_m(z')-E_n(z')|>0$ (not necessarily uniform in $n$)
\cite{2}. What happens when these conditions for some $n$
are violated in certain 
submanifolds of the slow phase space is an interesting question\footnote{In molecular
physics this leads to a rearrangement of the orbital binding sructure of atoms
and can be observed e.g. in organic molecules where the process is called 
isomerization.} 
which has 
to be considered in a case by case analysis.\\
\\
The next building blocks of the SAPT scheme are the spectral projection 
symbols
\be \label{2.1.7}   
\pi_{n,0}(z'):=\sum_{a=1}^{d_n}\; e_{n,a}(z')\; <e_{n,a}(z'),.>_{{\cal H}_f}
\ee
and the unitary symbols 
\be \label{2.1.8}
u_{n,0}(z'):=\sum_{n\in \mathbb{N}}\;\sum_{a=1}^{d_n}\;
b_{n,a}\; <e_{n,a}(z'),.>_{{\cal H}_f}
\ee
where $b_{n,a}$ is any orthormal basis of ${\cal H}_f$. 
The important point is that the {\it reference vectors} 
$b_{n,a}$ do not depend on $z'$ and we can 
pick them conveniently according to the concrete physical application,
e.g. $b_{n,a}:=e_{n,a}(z'=0)$. With their help we can define the 
{\it reference projection}
\be \label{2.1.9}
r_n:=\sum_{a=1}^{d_n}\; b_{n,a}\; <b_{n,a},.>_{{\cal H}_f}
\ee

The technical 
relevance of this {\it reference structure} is that $r_n$ in contrast 
to $\pi_{n,0}$ does not receive adiabatic corrections throughout 
the application of the 
SAPT scheme and thus always defines an {\it exact projector}
\be \label{2.1.10}
R_n=W(r_n)=r_n \otimes 1_s
\ee
which will be crucial for the adiabatic expansion and its spectral analysis. 
Note also that 
$u_{n,0}=u_0$ is in fact independent of $n$ while its adiabatic corrections 
will depend on $n$. In general we have series expansions 
\ba \label{2.1.11}
&& \pi_n(z')=\lim_{N\to \infty}\; \pi_{n,N}(z'),\;
\pi_{n,N}(z')=\sum_{k=0}^N\; \epsilon^k\; \pi_n^{(k)}(z'),\;
\pi_n^{(0)}(z')=\pi_{n,0}(z')
\nonumber\\
&& u_n(z')=\lim_{N\to \infty}\; u_{n,N}(z'),\;
u_{n,N}(z')=\sum_{k=0}^N\; \epsilon^k\; u_n^{(k)}(z'),\;
u_n^{(0)}(z')=u_{n,0}(z')
\ea
where the $\pi_{n,N},\; u_{n,N}$ are approximate Moyal 
projections and Moyal unitarities
up to corrections of order $\epsilon^{N+1}$, see below. Due to the 
linearity of the Weyl map we have 
\ba \label{2.1.11a}
\Pi_n&:=&W(\pi_n),\;
\Pi_{n,N}:=W(\pi_{n,N}),\;
\Pi^{(k)}_n:=W(\pi^{(k)}_n),\;
\nonumber\\
U_n&:=&W(u_n),\;
U_{n,N}:=W(u_{n,N}),\;
U^{(k)}_n:=W(u^{(k)}_n)
\nonumber\\
\Pi_n &=&
\lim_{N\to \infty} \Pi_{n,N},\;\Pi_{n,N}=\sum_{k=0}^N\;\epsilon^k\; \Pi_n^{(k)}
\nonumber\\
U_n &=&
\lim_{N\to \infty} U_{n,N},\;U_{n,N}=\sum_{k=0}^N\;\epsilon^k\; U_n^{(k)}
\ea
As we will spell out in detail below, the operators 
$\Pi_{n,N}:=W(\pi_{n,N}),\;U_{n,N}:=W(u_{n,N})$ will be  
constructed in such a way that the operator 
$U_{n,N}\; H\; U_{n,N}^\dagger,\; H=W(h)$ 
preserves the subspace $R_n {\cal H}$
up to corrections of order $\epsilon^{N+1}$. It thus coincides there 
up corrections of order $\epsilon^{N+1}$ with the operator 
$H_{n,N}=R_n U_{n,N} H U_{n,N}^\dagger R_n$ on the Hilbert subspace
$R_n {\cal H}\cong \mathbb{C}^{d_n}\otimes {\cal H}_s$. The importance 
of the reference structure now becomes manifest: While we have 
$H_{n,N}=U_{n,N} \tilde{H}_{n,N} U_{n,N}^\dagger$ up to corrections of 
order $\epsilon^{N+1}$ where $\tilde{H}_{n,}=\Pi_{n,N} H \Pi_{n,N}$,
since $\Pi_{n,N}$ is not an exact projector, the perhaps more 
natural operator 
$\tilde{H}_{n,N}$ which would not involve the objects $U_{n,N}$
does not preserve the subspace $\Pi_{n,N} {\cal H}$.
Hence while the spectral analysis of $H_{n,N}$ on $R_n {\cal H}$ can 
be preformed in the standard way, it would be unclear how to do 
that for $\tilde{H}_{n,N}$ on $\Pi_{n,N} {\cal H}$. Note that the 
problem is worse than $\tilde{H}_{n,N}$ not preserving its domain 
{\it within} $\Pi_{n,N}{|cal H}$, it actually maps its domain {\it outside} 
of $\Pi_{n,N}{\cal H}$ .   

This is then the perturbative  
{\it adiabatic decoupling} that we wanted to achieve. The spectrum 
of $H_{n,N}$, denoted by $E_{n,N}$  is referrred to as the $n$-th 
{\it energy band}. If $f_{n,N}\in R_n {\cal H}$ is a generalised eigenvector 
of $H_{n,N}$ with eigenvalue $\lambda$ 
then up to corrections of order $\epsilon^{N+1}$ the vector 
$\tilde{f}_{n,N}=U_{n,N}^\dagger f_{n,N}$ is a generalisd eigenvector 
of $H$ with the same eigenvalue since (we drop the $O(\epsilon^{N+1})$
terms)  
\ba \label{2.1.12}
&& H \tilde{f}_{n,N}=H U_{n,N}^\dagger R_n U_{n,N} U_{n,N}^\dagger f_{n,N}=
H \Pi_{n,N} U_{n,N}^\dagger f_{n,N}=
\Pi_{n,N} H U_{n,N}^\dagger f_{n,N}
\nonumber\\
&=& U_{n,N}^\dagger (R_n U_{n,N}  H U_{n,N}^\dagger R_n) f_{n,N}=
U_{n,N}^\dagger H_{n,N} f_{n,N}=
\lambda \tilde{f}_{n,N}
\ea
The approximate eigenvector $\tilde{f}_{n,N}$ is an element of the 
approximately invariant subspace $\Pi_{n,N} {\cal H}$ up to 
corrections of order $O(\epsilon^{N+1})$ because (we again drop the 
corrections) 
\be \label{2.1.13}
\tilde{f}_{n,N}=U_{n,N}^\dagger R_n U_{n,N} U_{n,N}^\dagger f_{n,N}
=\Pi_{n,N} \tilde{f}_{n,N}
\ee
In this way the $U_{n,N}$ are displayed as an auxiliary structure 
introduced in order to solve the spectral problem including backreation
but they have no further fundamental relevance as is also clear from the 
fact that they are not uniquely determined by the perturbative 
scheme in contrast to the $\Pi_{n,N}$. In particular, the $U_{n,N}$ are 
not to be confused with the unitary map $V$ that maps $\cal H$ to 
$L_2(\sigma(H),d\mu)$, granted to exist by the spectral theorem, where
$\sigma(H)$ is the spectrum of $H$ and $\mu$ its spectral measure
(in that Hilbert space, $H$ is a multiplication operator). This is 
already clear from the fact that $U_{n,N}$ generically depends on $n$
while $V$ does not.      

The fact that the $\Pi_{n,N}$ approximately commute with $H$ and 
are approximate projections displays them as approximants of spectral 
projections of $H$ on the part $E_{n,N}$ of the spectrum. 
The spectral projections are of course not necessarily 
mutually orthogonal even if the gap condition holds (for instance,
$h(z')$ could have pure point spectrum but $H$ could have absolutely
continuous spectrum), unless the energy bands are mutually disjoint.\\
\\
We proceed to detail the perturbative analysis.

\subsection{Perturbative Construction}
\label{s2.2}

The objective of the construction is to compute the set of operators 
\be \label{2.i}
\Pi_n:=W(\pi_n),\; U_n:=W(u_n),\; H:=W(h),\; R_n=W(r_n)=r_n\otimes 1_s
\ee
such that the following relations hold:
\be \label{2.ii}
\Pi_n^2-\Pi_n=\Pi_n^\dagger-\Pi_n=[\Pi_n,H]=
U_n U_n^\dagger-1_{{\cal H}}=
U_n^\dagger U_n -1_{{\cal H}}=U_n \Pi_n U_n^\dagger-R_n=0
\ee
Here $(.)^\dagger$ denotes the adjoint on ${\cal H}_f\otimes {\cal H}_s$.
Using the properties of Weyl quantisation this will be granted by 
the corresponding symbol relations 
\be \label{2.iii}
\pi_n\ast \pi_n-\pi_n=\pi_n^\dagger -\pi_n=\pi_n\ast h-h\ast \pi_n=
u_n\ast u_n^\dagger-1_f=  
u_n^\dagger\ast u_n-1_f=u_n\ast \pi_n \ast u_n^\dagger-r_n=0
\ee  
Here $(.)^\dagger$ denotes the adjoint on ${\cal H}_f$ and it should 
be clear from the lower case and upper case letter notation which adjoint
is being taken.

To obtain the $\pi_n, u_n$ define $\pi_{n,0},\; u_{n,0}$ as in 
(\ref{2.1.7}), (\ref{2.1.8}) and expand as in (\ref{2.1.11}). 
The symbols $\pi^{(k)}_n,\; u_n^{(k)}$ for $k>0$ are now defined 
inductively by requiring the approximate relations 
\ba \label{2.iv}
&& \pi_{n,N}\ast \pi_{n,N}-\pi_{n,N}=\pi_{n,N}^\dagger-\pi_{n,N}=
\pi_{n,N}\ast h-h\ast \pi_{n,N}=
u_{n,N}\ast u_{n,N}^\dagger-1_f  
\nonumber\\
&=& u_{n,N}^\dagger\ast u_{n,N}-1_f=u_{n,N}\ast \pi_{n,N} 
\ast u_{n,N}^\dagger-r_n=O(\epsilon^{N+1})
\ea  
where $O(\epsilon^{N+1})$ is a symbol whose leading order 
in its $\epsilon$ expansion is $\epsilon^{N+1}$ or higher. 
Since for any two symbols $a,b$ we have $a\ast b=ab +O(\epsilon)$ 
we note that (\ref{2.iv}) is certainly satisfied for $N=0$ since 
\be \label{2.v}
\pi_{n,0}\pi_{n,N}-\pi_{n,0}=
\pi_{n,0}^\dagger-\pi_{n,0}=
[\pi_{n,0},h_0]=
u_{n,0} u_{n,0}^\dagger-1_f=  
u_{n,0}^\dagger\ast u_{n,0}-1_f=u_{n,0} \pi_{n,0} 
\ast u_{n,0}^\dagger-r_n=0
\ee  
and we used $[a,h]=[a,h_0]+O(\epsilon)$ for any symbol $a$. Actually,
the construction only grants that in this way we obtain (\ref{2.iii}) 
but with zero replaced by $O(\epsilon^\infty)$ (e.g. $e^{-1/\epsilon}$). 
In \cite{2} resolvent methods are used to actually substitute 
$O(\epsilon^\infty)$ exactly by zero but for our purposes finite order 
approximations will be sufficient.

Also note that we do not impose any conditions on $O(\epsilon^{N+1})$.
In the ideal scenario we would like it to be of the form 
$\epsilon^{N+1} f(z')$ where 
$f(z')$ is a bounded operator on ${\cal H}_f$ with bound $||f(z')||$ 
perhaps even uniform on the slow phase space. We refer to \cite{2}
for circumstances under which one gets such results, however, these do
not apply here and we confine ourselves to a formal power expansion.\\
\\
We will now inductively construct first $\pi_{n,N}$ and after that $u_{n,N}$.

\subsubsection{Construction of the Moyal projections}
\label{2.2.1}

We will see that it is possible to construct all $\pi_n^{(k)}$ such that 
$(\pi_n^{(k)})^\dagger=\pi_n^{(k)}$ exactly for all $k$ thus implying
\be \label{2.viii})
\pi_{n,N}^\dagger=\pi_{n,N},\; \Pi_{n,N}^\dagger=\Pi_{n,N}
\ee
exactly for all $N$ which is included in our set of induction assumptions.

We isolate the leading order contributions
\be \label{2.xiii}
\pi_{n,N}\ast \pi_{n,N}-\pi_{n,N}=:a_{n,N} \; \epsilon^{N+1}+O(\epsilon^{N+2}),
\;\;
\pi_{n,N}\ast h-h\ast \pi_{n,N}=:b_{n,N}\; \epsilon^{N+1}+O(\epsilon^{N+2})
\ee
The symbols $a_{n,N}, b_{n,N}$ are symmetric and antisymmetric operators
on ${\cal H}_f$ respectively due to the Moyal identity for symbols $f,g$
\be \label{2.xiiia}
(f\ast g)^\dagger=g^\dagger\ast f^\dagger
\ee
We obtain 
\ba \label{2.xiv}
&& \pi_{n,N+1}\ast \pi_{n,N+1}-\pi_{n,N+1}
\nonumber\\
&=&
\pi_{n,N}\ast \pi_{n,N}-\pi_{n,N}
+\epsilon^{N+1}\{[\pi_n^{(N+1)}\ast \pi_{n,N}+
\pi_{n,N}\ast \pi_n^{(N+1)}\ast-\pi_n^{(N+1)}\}+\epsilon^{2(N+1)}
\pi_n^{(N+1)}\ast\pi_n^{(N+1)}
\nonumber\\
&=&\epsilon^{N+1}\{a_{n,N}+\pi_n^{(N+1)} \pi_n^{(0)}+
\pi_n^{(0)}\pi_n^{(N+1)}-\pi_n^{(N+1)}\}+O(\epsilon^{(N+2)}
\ea
We conclude
\be \label{2.xv}
-a_{n,N}=\pi_n^{(N+1)} \pi_n^{(0)}+
\pi_n^{(0)}\pi_n^{(N+1)}-\pi_n^{(N+1)}
\ee
Since we will it frequently let us abbreviate
\be \label{2.xvi}
P_n=\pi_{n,0},\;P_n^\perp=1_{{\cal H}_f}-P_n
\ee
Then by projecting (\ref{2.xv}) to the block diagonal pieces
\be \label{2.xvii}   
-P_n a_{n,N} P_n=P_n \pi_n^{(N+1)} P_n,\;\;
P_n^\perp a_{n,N} P_n^\perp=P_n^\perp \pi_n^{(N+1)} P_n^\perp
\ee
while for the projection to the off block diagonal pieces we obtain
the consistency condition
\be \label{2.xviii}
P_n a_{n,N} P_n^\perp=
P_n^\perp a_{n,N} P_n=0
\ee
This identity follows from the defining equation (\ref{2.xiii})
which we project to the off block diagonal pieces
\ba \label{2.xix}
&& \epsilon^{N+1}\;P_n a_{n,N} P_n^\perp=
P_n(\pi_{n,N}\ast \pi_{n,N}-\pi_{n,N})P_n^\perp+O(\epsilon^{N+2})
\nonumber\\
&=&
\pi_{n,N}(\pi_{n,N}\ast \pi_{n,N}-\pi_{n,N})(1-\pi_{n,N})+O(\epsilon^{N+2})
\nonumber\\
&=&
\pi_{n,N}\ast(\pi_{n,N}\ast \pi_{n,N}-\pi_{n,N})\ast(1-\pi_{n,N})
+O(\epsilon^{N+2})
\nonumber\\
&=&
(\pi_{n,N}\ast \pi_{n,N}-\pi_{n,N})\ast\pi_{n,N}\ast(1-\pi_{n,N})
+O(\epsilon^{N+2})
\nonumber\\
&=&-\epsilon^{2(N+1)} a_{n,N}\ast a_{n,N}+O(\epsilon^{N+2)}=O(\epsilon^{N+2})
\ea
where associativity of the Moyal product was used.

Accordingly, the off diagonal pieces of $\pi_n^{(N+1)}$ are still
undetermined. We use the second condition in (\ref{2.iv}) to fix
it. With the notation $[f,g]_\ast=f\ast g -g \ast f$ we have 
\ba \label{2.xx}
{[}\pi_{n,N+1},h]_\ast 
&=&
{[}\pi_{n,N},h]_\ast +
\epsilon^{N+1}\;[\pi_n^{N+1},h]_\ast
\nonumber\\
&=&
{[}\pi_{n,N},h]_\ast +
\epsilon^{N+1}\;[\pi_n^{N+1},h]_\ast
\nonumber\\
&=&
\epsilon^{N+1}\;\{b_{n,N}+[\pi_n^{N+1},\hat{h}_0]_\ast\} +O(\epsilon^{N+2})
\nonumber\\
&=&
\epsilon^{N+1}\;\{b_{n,N}+[\pi_n^{N+1},\hat{h}_0]\} +O(\epsilon^{N+2})
\ea
Accordingly
\be \label{2.xxi}
b_{n,N}+[\pi_n^{(N+1)},\hat{h}_0]=0
\ee
Projecting to the block diagonal pieces we obtain the consistency conditions 
using $[P_n,\hat{h}_0]=0=[P_n^\perp,\hat{h}_0]$
\ba \label{2.xxii}
0 &=& P_n b_{n,N} P_n+[P_n \pi_n^{(N+1)} P_n,\hat{h}_0]
= P_n b_{n,N} P_n-[P_n a_{n,N} P_n,\hat{h}_0]
\nonumber\\
0 &=& P_n^\perp b_{n,N} P_n^\perp+[P_n^\perp \pi_n^{(N+1)} 
P_n^\perp,\hat{h}_0]
= P_n^\perp b_{n,N} P_n^\perp+[P_n^\perp a_{n,N} P_n^\perp,\hat{h}_0]
\ea
Indeed, again using the defining equations (\ref{2.xiii})
\ba \label{2.xxiia}
&& \{P_n b_{n,N} P_n-[P_n a_{n,N} P_n,\hat{h}_0]\}\epsilon^{N+1}
= \{P_n [\pi_{n,N},h]_\ast P_n-
[P_n (\pi_{n,N}\ast\pi_{n,N}-\pi_{n,N}) P_n,\hat{h}_0]\}
+O(\epsilon^{N+2})
\nonumber\\
&=&
P_n\{[\pi_{n,N},h]_\ast-
[(\pi_{n,N}\ast\pi_{n,N}-\pi_{n,N}),\hat{h}_0]\}P_n
+O(\epsilon^{N+2})
\nonumber\\
&=&
P_n\{[\pi_{n,N},h]_\ast -
[(\pi_{n,N}\ast\pi_{n,N}-\pi_{n,N}),\hat{h}_0]_\ast\}P_n
+O(\epsilon^{N+2})
\nonumber\\
&=&
P_n\{[\pi_{n,N},h]_\ast -
[(\pi_{n,N}\ast\pi_{n,N}-\pi_{n,N}),h]_\ast\}P_n
+O(\epsilon^{N+2})
\nonumber\\
&=&
P_n\{2[\pi_{n,N},h]_\ast -
[\pi_{n,N}\ast\pi_{n,N},h]_\ast\}P_n
+O(\epsilon^{N+2})
\nonumber\\
&=&
P_n\{[(2\pi_{n,N}-\pi_{n,N}\ast\pi_{n,N}),h]_\ast\}P_n
+O(\epsilon^{N+2})
\nonumber\\
&=&
\pi_{n,N}\ast\{[(2\pi_{n,N}-\pi_{n,N}\ast\pi_{n,N}),h]_\ast\}\ast \pi_{n,N}
+O(\epsilon^{N+2})
\nonumber\\
&=&
2\pi_{n,N}\ast[\pi_{n,N},h]_\ast \ast \pi_{n,N}
-\pi_{n,N}\ast\pi_{n,N}[\pi_{n,N},h]_\ast \ast \pi_{n,N}
-\pi_{n,N}\ast[\pi_{n,N},h]_\ast \ast \pi_{n,N}\ast\pi_{n,N}
+O(\epsilon^{N+2})
\nonumber\\
&=&
(\pi_{n,N}-\pi_{n,N})\ast[\pi_{n,N},h]_\ast \ast \pi_{n,N}
+\pi_{n,N}\ast[\pi_{n,N},h]_\ast \ast(\pi_{n,N}-\pi_{n,N}\ast\pi_{n,N})
+O(\epsilon^{N+2})
\nonumber\\
&=& O(\epsilon^{N+2})
\ea
Similarly
\ba \label{2.xxiii}
&& \{P_n^\perp b_{n,N} P_n^\perp+[P_n^\perp a_{n,N} P_n^\perp,\hat{h}_0]\}
\epsilon^{N+1}
= \{P_n^\perp [\pi_{n,N},h]_\ast P_n^\perp+
[P_n^\perp (\pi_{n,N}\ast\pi_{n,N}-\pi_{n,N}) P_n^\perp,\hat{h}_0]\}
+O(\epsilon^{N+2})
\nonumber\\
&=&
P_n^\perp\{[\pi_{n,N},h]_\ast+
[(\pi_{n,N}\ast\pi_{n,N}-\pi_{n,N}),\hat{h}_0]\}P_n^\perp
+O(\epsilon^{N+2})
\nonumber\\
&=&
P_n^\perp\{[\pi_{n,N},h]_\ast+
[(\pi_{n,N}\ast\pi_{n,N}-\pi_{n,N}),\hat{h}_0]_\ast\}P_n^\perp
+O(\epsilon^{N+2})
\nonumber\\
&=&
P_n^\perp\{[\pi_{n,N},h]_\ast +
[(\pi_{n,N}\ast\pi_{n,N}-\pi_{n,N}),h]_\ast\}P_n^\perp
+O(\epsilon^{N+2})
\nonumber\\
&=&
P_n^\perp[\pi_{n,N}\ast\pi_{n,N},h]_\ast\}P_n^\perp
+O(\epsilon^{N+2})
\nonumber\\
&=&
\pi_{n,N}^\perp\ast [\pi_{n,N}\ast\pi_{n,N}),h]_\ast \ast \pi_{n,N}\perp
+O(\epsilon^{N+2})
\nonumber\\
&=&
\pi_{n,N}^\perp \ast\pi_{n,N}[\pi_{n,N},h]_\ast \ast \pi_{n,N}^\perp
+\pi_{n,N}^\perp \ast[\pi_{n,N},h]_\ast \ast \pi_{n,N}\ast\pi_{n,N}^\perp
+O(\epsilon^{N+2})
\nonumber\\
&=& O(\epsilon^{N+2})
\ea
Thus indeed (\ref{2.xxi}) is only a condition on the block off diagonal 
projections
\be \label{2.xxiv}
P_n b_{n,N} P_n^\perp+[P_n \pi_n^{(N+1)} P_n^\perp,\hat{h}_0]=0
=P_n^\perp b_{n,N} P_n+[P_n^\perp \pi_n^{(N+1)} P_n,\hat{h}_0]
\ee
We only use the first condition of (\ref{2.xxiv}) as the second follows by
taking the adjoint of the first if we manage to keep symmetry 
of $\pi_n^{(N+1)}$. We have with 
$f_{n,N+1}:=P_n \pi_n^{(N+1)} P_n^\perp$ using $P_n^2=P_n,\;[P_n,\hat{h}_0]=0$
\ba \label{2.xxv} 
&& [f_{n,N+1},\hat{h}_0]=
f_{n,N+1}(P_n^\perp \hat{h}_0 P_n^\perp)
-(P_n \hat{h}_0 P_n)f_{n,N+1}
=f_{n,N+1}\;(\sum_{m\not=n} E_m P_m)-E_n \;c_{n,N+1}\; P_n^\perp
\nonumber\\
&=& f_{n,N+1} (\sum_{m\not=n} (E_m-E_n)  P_m)
=:f_{n,N+1} \hat{h}_{0n}^\perp
\ea
The operator $\hat{h}_{0n}^\perp$ has the inverse 
on $P_n^\perp {\cal H}_f$ given by 
\be \label{2.xxvi}
\Delta_n:=\sum_{m\not=n}\; (E_m-E_n)^{-1} \; P_m
\ee
provided the {\it no band crossing condition} $E_m(z)-E_n(z)\not=0$ 
for all $z,\;m\not=n$ holds. As mentioned 
before the strongest condition would be that
gap number 
\be \label{2.xxvii}
g_n:=\sup_{z,m\not= n} |E_m(z)-E_n(z)| 
\ee
should be positive but we will not rely on this in our purely formal
investigation. Thus we find 
\be \label{2.xxviii}
f_{n,N+1}=-P_n b_{n,N} P_n^\perp \Delta_n
\ee
Note that both the l.h.s. and the r.h.s. of (\ref{2.xxviii}) annihilate 
$P_n {\cal H}_f$.\\
\\
Collecting all terms we thus have computed
\be \label{2.xxix}
\pi_n^{(N+1)}=-P_n a_{n,N} P_n+
P_n^\perp a_{n,N} P_n^\perp-P_n b_{n,N} P_n^\perp \Delta_n
+\Delta_n P_n^\perp b_{n,N} P_n
\ee
from $a_{n,N},\;b_{n,N}$ given in (\ref{2.xiii}). One easily checks 
that (\ref{2.xxix}) is symmetric.

\subsubsection{Construction of the Moyal unitarities}
\label{s2.2.2}

We now turn to constructing $u_{n,N+1}$ from $u_{n,N},\pi_{n,N+1}$.
We isolate the leading orders 
\ba \label{2.xxxiv}
&& u_{n,N}\ast u_{n,N}^\dagger-1_f=:\epsilon^{N+1} c_{n,N}+ O(\epsilon^{N+2}),\; 
u_{n,N}^\dagger\ast u_{n,N}-1_f=:\epsilon^{N+1} e_{n,N}+ O(\epsilon^{N+2}),\; 
\nonumber\\
&& u_{n,N}\ast \pi_{n,N} \ast u_{n,N}^\dagger-r_n=:\epsilon^{N+1} d_{n,N}+
O(\epsilon^{N+2})
\ea
with $c_{n,N}, d_{n,N},e_{n,N}$ symmetric and find 
\ba \label{2.xxxv}
&&u_{n,N+1}\ast u_{n,N+1}^\dagger-1_f
=[u_{n,N}\ast u_{n,N}^\dagger-1_f]+\epsilon^{N+1}
\{u_n^{(N+1)}\ast u_{n,N}^\dagger+u_{n,N}\ast (u_n^{(N+1)})^\dagger\}
\nonumber\\
&& +\epsilon^{2(N+1)} u_n^{(N+1)}\ast (u_n^{(N+1)})^\dagger
\nonumber\\
&=& \{c_{n,N}+u_n^{(N+1)}\ast u_{n,0}^\dagger+u_{n,0}\ast 
(u_n^{(N+1)})^\dagger\}\epsilon^{N+1}+O(\epsilon^{N+2})
\nonumber\\
&=& \{c_{n,N}+u_n^{(N+1)} u_{n,0}^\dagger+u_{n,0} 
(u_n^{(N+1)})^\dagger\}\epsilon^{N+1}+O(\epsilon^{N+2})
\ea
Similarly
\be \label{2.xxxvi}
u_{n,N+1}^\dagger\ast u_{n,N+1}-1_f
=\{e_{n,N}+(u_n^{(N+1)})^\dagger u_{n,0}+u_{n,0}^\dagger 
u_n^{(N+1)}\}\epsilon^{N+1}+O(\epsilon^{N+2})
\ee
and 
\be \label{2.xxxvii}
u_{n,N+1}\ast \pi_{n,N+1} \ast u_{n,N+1}^\dagger-r_n=
\epsilon^{N+1}\{d_{n,N}+u_{n,0} \pi_n^{(N+1)} u_{n,0}^\dagger
+u_n^{(N+1)} \pi_{n,0} u_{n,0}^\dagger
u_{n,0} \pi_{n,0} (u_n^{(N+1)})^\dagger\}
+
O(\epsilon^{N+2})
\ee   
Accordingly
\ba \label{2.xxxviii}
0 &=& c_{n,N}+u_n^{(N+1)} u_{n,0}^\dagger+u_{n,0} 
(u_n^{(N+1)})^\dagger
\nonumber\\
0= e_{n,N}+(u_n^{(N+1)})^\dagger u_{n,0}+u_{n,0}^\dagger 
u_n^{(N+1)}
\nonumber\\
0 &=& d_{n,N}+u_{n,0} \pi_n^{(N+1)} u_{n,0}^\dagger
+u_n^{(N+1)} \pi_{n,0} u_{n,0}^\dagger
u_{n,0} \pi_{n,0} (u_n^{(N+1)})^\dagger
\ea
We isolate $(u_n^{(N+1)})^\dagger$ from the first two equations
\be \label{2.xxxix}
(u_n^{(N+1)})^\dagger=-u_{n,0}^\dagger(c_{n,N}+u_n^{(N+1)} u_{n,0}^\dagger)
=-(e_{n,N}+u_{n,0}^\dagger u_n^{(N+1)}) u_{n,0}^\dagger
\ee
implying the identity 
\be \label{2.xl}
e_{n,N} u_{n,0}^\dagger-u_{n,0}^\dagger c_{n,N}=0
\ee
This is identically satisfied by induction assumption since 
\ba \label{2.xli}
&& 
(e_{n,N} u_{n,0}^\dagger-u_{n,0}^\dagger c_{n,N})\epsilon^{N+1}
=(u_{n,N}^\dagger \ast u_{n,N}-1_f)u_{n,0}^\dagger-
u_{n,0}^\dagger (u_{n,N}\ast u_{n,N}^\dagger-1_f)+O(\epsilon^{N+2})
\nonumber\\
&=& 
(u_{n,N}^\dagger \ast u_{n,N}-1_f)\ast u_{n,0}^\dagger-
u_{n,0}^\dagger\ast (u_{n,N}\ast u_{n,N}^\dagger-1_f)+O(\epsilon^{N+2})
\nonumber\\
&=& 
(u_{n,N}^\dagger \ast u_{n,N}-1_f)\ast u_{n,N}^\dagger-
u_{n,N}^\dagger\ast (u_{n,N}\ast u_{n,N}^\dagger-1_f)+O(\epsilon^{N+2})
\nonumber\\
&=& O(\epsilon^{N+2})
\ea
Accordingly we may use henceforth 
\be \label{2.xlii}
e_{n,N}=u_{n,0}^\dagger c_{n,N} u_{n,0}^\dagger
\ee
Substituting (\ref{2.xxxix}) into the third relation of (\ref{2.xxxviii})
yields
\ba  \label{2.xliii}
&& -(d_{n,N}+u_{n,0} \pi_n^{(N+1)} u_{n,0}^\dagger) =
u_n^{(N+1)} \pi_{n,0} u_{n,0}^\dagger
-u_{n,0} \pi_{n,0} 
u_{n,0}^\dagger(c_{n,N}+u_n^{(N+1)} u_{n,0}^\dagger)
\nonumber\\
&=&
u_n^{(N+1)} u_{n,0}^\dagger r_n
-r_n (c_{n,N}+u_n^{(N+1)} u_{n,0}^\dagger)
\nonumber\\
&=& - r_n c_{n,N}+[u_n^{(N+1)} u_{n,0}^\dagger,r_n]
\ea
or 
\be \label{2.xliv}
[r_n, u_n^{(N+1)} u_{n,0}^\dagger]=
 d_{n,N}+u_{n,0} \pi_n^{(N+1)} u_{n,0}^\dagger-r_n c_{n,N}
\ee
Projecting with $r_n$ or $r_n^\perp=1_f-r_n$ from both sides gives the 
identities 
\be \label{2.xlv}
0=r_n (d_{n,N}+u_{n,0} \pi_n^{(N+1)} u_{n,0}^\dagger-r_n c_{n,N}) r_n
=r_n^\perp (d_{n,N}+u_{n,0} \pi_n^{(N+1)} u_{n,0}^\dagger-r_n c_{n,N}) 
r_n^\perp
\ee
which are again identically satisfied by induction assumption since,
remembering (\ref{2.xxix}) which we write in the form (and using 
$P_n^\perp \Delta_n=\Delta_n P_n^\perp=\Delta_n$)
\be \label{2.xlvi}
u_{n,0}\pi_n^{(N+1)} u_{n,0}^\dagger=-r_n u_{n,0} a_{n,N} u_{n,0}^\dagger
r_n+
r_n^\perp u_{n,0} a_{n,N} u_{n,0}^\dagger
r_n^\perp-r_n u_{n,0} b_{n,N} \Delta_n u_{n,0}^\dagger r_n^\perp
+r_n^\perp u_{n,0} \Delta_n  b_{n,N} u_{n,0}^\dagger r_n
\ee
whence
\ba \label{2.xlvii} 
&& \epsilon^{N+1}\;
r_n (d_{n,N}+u_{n,0} \pi_n^{(N+1)} u_{n,0}^\dagger-r_n c_{n,N}) r_n
=
\epsilon^{N+1}\;
r_n (d_{n,N}- u_{n,0} a_{n,N} u_{n,0}^\dagger -c_{n,N}) r_n
\nonumber\\
&=&
r_n ([u_{n,N}\ast \pi_{n,N} \ast u_{n,N}^\dagger-r_n]
- u_{n,0} [\pi_{n,N}\ast \pi_{n,N}-\pi_{n,N}] u_{n,0}^\dagger 
-[u_{n,N}\ast u_{n,N}^\dagger-1_f]) r_n
+O(\epsilon^{N+2})
\nonumber\\
&=&
r_n (u_{n,N}\ast \pi_{n,N} \ast u_{n,N}^\dagger
- u_{n,0}\ast (\pi_{n,N}\ast \pi_{n,N}-\pi_{n,N})\ast u_{n,0}^\dagger 
-u_{n,N}\ast u_{n,N}^\dagger) r_n
+O(\epsilon^{N+2})
\nonumber\\
&=&
r_n u_{n,N}\ast (\pi_{n,N} 
-\pi_{n,N}\ast \pi_{n,N}+ \pi_{n,N} 
-1_f) \ast u_{n,N}^\dagger r_n
+O(\epsilon^{N+2})
\nonumber\\
&=&-
r_n u_{n,N}\ast (\pi_{n,N}-1_f)\ast (\pi_{n,N}-1_f)\ast 
u_{n,N}^\dagger r_n
+O(\epsilon^{N+2})
\ea
We abbreviate the following $O(\epsilon^{N+1})$ objects
\be \label{2.xlviii}
D_n:=u_{n,N}\ast \pi_{n,N}\ast u_{n,N}^\dagger-r_n,\;
C_n:=u_{n,N}\ast u_{n,N}^\dagger-1_f,\;
E_n:=u_{n,N}^\dagger\ast u_{n,N}-1_f,\;
A_n:=\pi_{n,N}\ast\pi_{n,N}-\pi_{n,N}
\ee
any bilinear combination of which is thus of order $\epsilon^{N+2}$ and 
continue (\ref{2.xlvii}) (in the first step, use that $r_n$ is a constant
on the slow phase space)
\ba \label{2.xlix}
&&
r_n u_{n,N}\ast (\pi_{n,N}-1_f)\ast (\pi_{n,N}-1_f)\ast 
u_{n,N}^\dagger r_n
+O(\epsilon^{N+2})
\nonumber\\
&=&
r_n \ast u_{n,N}\ast (\pi_{n,N}-1_f)\ast (\pi_{n,N}-1_f)\ast 
u_{n,N}^\dagger \ast r_n
+O(\epsilon^{N+2})
\nonumber\\
&=& (u_{n,N}\ast\pi_{n,N}\ast u_{n,N}^\dagger-D_n)\ast 
u_{n,N}\ast (\pi_{n,N}-1_f)\ast (\pi_{n,N}-1_f)\ast 
u_{n,N}^\dagger \ast (u_{n,N} \ast \pi_{n,N} \ast u_{n,N}^\dagger -D_n)
+O(\epsilon^{N+2})
\nonumber\\
&=& u_{n,N}\ast\pi_{n,N}\ast (u_{n,N}^\dagger \ast 
u_{n,N})\ast (\pi_{n,N}-1_f)\ast (\pi_{n,N}-1_f)\ast 
(u_{n,N}^\dagger \ast u_{n,N}) \ast \pi_{n,N} \ast u_{n,N}^\dagger
\nonumber\\
&& - D_n\ast 
u_{n,N}\ast (\pi_{n,N}-1_f)\ast (\pi_{n,N}-1_f)\ast 
(u_{n,N}^\dagger \ast (u_{n,N}) \ast \pi_{n,N} \ast u_{n,N}^\dagger
\nonumber\\
&=& -u_{n,N}\ast\pi_{n,N}\ast (u_{n,N}^\dagger \ast 
u_{n,N})\ast (\pi_{n,N}-1_f)\ast (\pi_{n,N}-1_f)\ast 
u_{n,N}^\dagger \ast (u_{n,N} \ast \pi_{n,N} \ast D_n
+O(\epsilon^{N+2})
\nonumber\\
&=& u_{n,N}\ast\pi_{n,N}\ast (E_n+1_f) \ast (\pi_{n,N}-1_f)\ast 
(\pi_{n,N}-1_f)\ast 
(E_n+1_f) \ast \pi_{n,N} \ast u_{n,N}^\dagger
\nonumber\\
&& - D_n\ast 
u_{n,N}\ast (\pi_{n,N}-1_f)\ast (\pi_{n,N}-1_f)\ast 
(E_n+1_f) \ast \pi_{n,N} \ast u_{n,N}^\dagger
\nonumber\\
&=& -u_{n,N}\ast\pi_{n,N}\ast (E_n+1_f)\ast (\pi_{n,N}-1_f)\ast 
(\pi_{n,N}-1_f)\ast 
u_{n,N}^\dagger \ast (u_{n,N} \ast \pi_{n,N} \ast D_n
+O(\epsilon^{N+2})
\nonumber\\
&=& u_{n,N}\ast\pi_{n,N}\ast E_n \ast (\pi_{n,N}-1_f)\ast 
A_n \ast u_{n,N}^\dagger
+
u_{n,N}\ast\pi_{n,N}\ast A_n \ast 
(\pi_{n,N}-1_f)\ast E_n\ast \pi_{n,N} \ast u_{n,N}^\dagger
\nonumber\\
&& +
u_{n,N}\ast A_n \ast A_n \ast u_{n,N}^\dagger
\nonumber\\
&& - D_n\ast 
u_{n,N}\ast (\pi_{n,N}-1_f)\ast \{
(\pi_{n,N}-1_f)\ast E_n \ast \pi_{n,N} +A_n\} 
\ast u_{n,N}^\dagger
\nonumber\\
&=& -u_{n,N}\ast
\{
\pi_{n,N}\ast E_n\ast (\pi_{n,N}-1_f)+A_n\}\ast (\pi_{n,N}-1_f)\ast 
u_{n,N}^\dagger \ast (u_{n,N} \ast \pi_{n,N} \ast D_n
+O(\epsilon^{N+2})
\nonumber\\
&=& O(\epsilon^{N+2})
\ea
The second identity in (\ref{2.xlv}) folllows similarly
\ba \label{2.l}
&& \epsilon^{N+1}
r_n^\perp (d_{n,N}+u_{n,0} \pi_n^{(N+1)} u_{n,0}^\dagger-r_n c_{n,N}) 
r_n^\perp
=\epsilon^{N+1}
r_n^\perp (d_{n,N}+u_{n,0} a_{n,N} u_{n,0}^\dagger) 
r_n^\perp
\nonumber\\
&=&
r_n^\perp (u_{n,N}\ast \pi_{n,N}\ast u_{n,N}^\dagger-r_n+
u_{n,0}(\pi_{n,N}\ast\pi_{n,N}-\pi_{n,N}) u_{n,0}^\dagger) 
r_n^\perp +O(\epsilon^{N+2})
\nonumber\\
&=&
r_n^\perp (u_{n,N}\ast \pi_{n,N}\ast u_{n,N}^\dagger+
u_{n,0}\ast(\pi_{n,N}\ast\pi_{n,N}-\pi_{n,N})\ast u_{n,0}^\dagger) 
r_n^\perp +O(\epsilon^{N+2})
\nonumber\\
&=&
r_n^\perp u_{n,N}\ast(\pi_{n,N}\ast+
\pi_{n,N}\ast\pi_{n,N}-\pi_{n,N})\ast u_{n,N}^\dagger 
r_n^\perp +O(\epsilon^{N+2})
\nonumber\\
&=&
r_n^\perp \ast u_{n,N}\ast 
\pi_{n,N}\ast\pi_{n,N}\ast u_{n,N}^\dagger \ast
r_n^\perp +O(\epsilon^{N+2})
\nonumber\\
&=&
(1_f-u_{n,N}\ast \pi_{n,N} \ast u_{n,N}^\dagger-D_n)\ast u_{n,N}\ast 
\pi_{n,N}\ast\pi_{n,N}\ast u_{n,N}^\dagger \ast
(1_f-u_{n,N}\ast \pi_{n,N} \ast u_{n,N}^\dagger-D_n) +O(\epsilon^{N+2})
\nonumber\\
&=&
(u_{n,N}\ast(1_f-\pi_{n,N} \ast u_{n,N}^\dagger\ast u_{n,N})-
D_n\ast u_{n,N})\ast 
\pi_{n,N}\ast\pi_{n,N}\ast
\nonumber\\
&& ((1_f-u_{n,N}^\dagger\ast u_{n,N}\ast \pi_{n,N}) \ast u_{n,N}^\dagger
-u_{n,N}^\dagger\ast D_n) +O(\epsilon^{N+2})
\nonumber\\
&=&
(u_{n,N}\ast(1_f-\pi_{n,N} \ast (1_f+E_n))-D_n\ast u_{n,N})\ast 
\pi_{n,N}\ast\pi_{n,N}\ast
\nonumber\\
&& ((1_f-(1_f+E_n)\ast \pi_{n,N}) \ast u_{n,N}^\dagger
-u_{n,N}^\dagger\ast D_n) +O(\epsilon^{N+2})
\nonumber\\
&=&
[u_{n,N}\ast A_n-(u_{n,N}\ast\pi_{n,N}\ast E_n+D_n\ast u_{n,N}) \ast \pi_{n,N}]
\ast\pi_{n,N}\ast
\nonumber\\
&& [A_n\ast u_{n,N}^\dagger
-\pi_{n,N}\ast(E_n\ast u_{n,N}^\dagger+u_{n,N}^\dagger \ast D_n]
+O(\epsilon^{N+2})
\nonumber\\
&=& O(\epsilon^{N+2})
\ea
where we again used constancy of $r_n$.

It follows that the block diagonal pieces of $u_n^{(N+1)}$ wrt $r_n$ remain
undetermined (we can choose them to vanish for simplicity). We thus project 
to the block off diagonal parts of (\ref{2.xliv}) using again (\ref{2.xlvi})
\ba \label{2.li}
&&
r_n u_n^{(N+1)} u_{n,0}^\dagger r_n^\perp
= r_n(d_{n,N}+u_{n,0} \pi_n^{(N+1)} u_{n,0}^\dagger-r_n c_{n,N})r_n^\perp
\nonumber\\
&=& 
r_n(d_{n,N}-u_{n,0} b_{n,N} \Delta_n u_{n,0}^\dagger-c_{n,N})r_n^\perp
\nonumber\\
&&
-r_n^\perp u_n^{(N+1)} u_{n,0}^\dagger r_n
= r_n^\perp (d_{n,N}+u_{n,0} \pi_n^{(N+1)} u_{n,0}^\dagger
-r_n c_{n,N}) r_n
\nonumber\\
&=& r_n^\perp (d_{n,N}+u_{n,0} \Delta_n  b_{n,N} u_{n,0}^\dagger)r_n
\ea
Accordingly, if we indeed assume the block off diagonal terms to vanish 
for simplicity since $u_{n,N}$ is just an auxiliary structure
we find explicitly
\ba \label{2.lii}
u_n^{(N+1)} &=&
\{
r_n(d_{n,N}-u_{n,0} b_{n,N} \Delta_n u_{n,0}^\dagger-c_{n,N})r_n^\perp
\nonumber\\
&&
-r_n^\perp (d_{n,N}+u_{n,0} \Delta_n  b_{n,N} u_{n,0}^\dagger)r_n\} u_{n,0}
\ea

\subsubsection{Effective Hamiltonian}
\label{s2.2.3}

This concludes the perturbative construction. Using equations
(\ref{2.xiii}), (\ref{2.xxix}), (\ref{2.xxxiv}), (\ref{2.xlii}) and 
(\ref{2.lii}) we can compute $\pi_{n,N},\; u_{n,N}$ up to any 
finite order $N$. As outlined in section \ref{s2.1} we now construct 
the {\it effective symbol}
\be \label{2.liii}
h_{n,N}=r_n(u_{n,N}\ast h\ast u_{n,N}^\dagger)r_n
\ee
and from this the {\it effective Hamiltonian}
\be \label{2.liv}
H_{n,N}=W(h_{n,N})=R_n U_{n,N} H U_{n,N}^\dagger R_n
\ee
that preserves the subspace $R_n {\cal H}$. That subspace carries 
the orthonormal basis $b_{n,a}\otimes s_\alpha,\; a=1,..,d_n$ and 
$s_\alpha$ denotes an ONB of ${\cal H}_s$. The spectrum $E_{n,N}$ 
of $H_{n,N}$ gives an approximation of order $\epsilon^{N+1}$ of the 
$n$-th energy band $E_n$ of $H$. The advantage of $H_{N,n}$ is that 
it is effectively an operator on the rather small Hilbert space 
$\mathbb{C}^{d_n}\otimes {\cal H}_s$ while backreaction effects between 
the slow and fast sector are taken care of to the given order of 
approximation.

\section{Challenges of cosmological SAPT}
\label{s4}

As mentioned in the introduction and as it also transpires from the previous
section, SAPT was designed for quantum theories with a finite number of 
degrees of freedom. When we try to generalise to quantum field theories
such as second order quantum cosmological perturbation theory, we
meet challenges because some of the assumptions of the quantum mechanical 
setting do not automatically transfer to the QFT case. Here 
the role of the fast degrees of freedom is played by the infinite number
of inhomogeneous perturbations while the slow ones are the homogeneous 
modes. We will show this below, for the time being it may be sufficient 
to say that when cutting off the number of inhomogeneous modes in an
intermediate step in order to be in the quantum mechanical setting of SAPT
we can consider the system as a ``gas'' with a finite number of particles 
and the homogeneous mode takes the role of the centre of mass degree
of freedom. Note that the volume of the ``gas'' is finite because 
our models, even when considering only the classical level,  
are only well defined when the spatial slices are compact.   

As we will see, these kind of problems originating from the 
infinite number of degrees of freedom can be circumvented using a 
canonical transformation of the total system 
involving both homogeneous and inhomogeneous modes
which is exact up top second 
order in the perturbations. Here we will borrow ideas developed for the 
hybrid approach to Lopp Quantum Cosmology (LQC) \cite{8}.
However, then one meets the next problem:
The mass squared functions that enter the quantum field perturbation 
get modified as a result of that canonical transformation and 
are generically neither positive definite nor are they polynomials in the slow 
(homogeneous) degrees of freedom, an example being the Mukhanov-Sasaki mass
squared term. The latter property is already met 
for the purely homogeneous contribution to the Hamiltonian constraint,
however, the degree of npn-polynomiality gets {\it much worsened}.
As there is some amount of freedom in the choice of that canonical 
transformation, the indefiniteness of the mass squared term could perhaps be 
avoided by exploiting that freedom, however at the present stage of research 
we are not sure that this is the case and thus prepare ourselves for the 
worst case scenario. 

As far as the first problem is concerned, we offer 
two kinds of 
solutions: Either one considers a further canonical transformation 
just of the slow sector to new variables in terms of which the masses 
are manifestly non negative and declares the phase space as defined 
in terms of the old variables to be restricted 
by the positivity of the masses. Or one has to switch off modes by hand 
for which the frequency becomes negative as suggested first in 
\cite{17}. The first solution has the advantage that one stays within the 
standard QFT framework of positive mass squared, however at the price
of making the Hamiltonian even less polynomial and by a possibly non-physical 
restriction of the phase space. The second solution has the 
advantage of not worsening the amount of non polynomiality and of not
modifying the phase space of the slow sector, however, 
the physical interpretation of the mode off switching remains obscure:
Typically this happens at the classical big bang singularity and in the 
limit of vanishing scale factor, all modes would need to be removed. 
One could speculate whether this presents a self-regulating effect of a new kind 
in the sense that the matter density 
and thus curvature automatically vanishes as we approach the 
singularity, thus eventually avoiding the big bang. 

As far as the second problem is concerned, in LQC methods for dealing 
with non-polynomial expressions in momentum variables but polynomial 
in configuration variables adapted from 
techniques developed for the full theory \cite{18} have been employed.
These become available because one uses a representation of the slow
sector of the theory which is not unitarily equivalent to the 
Schr\"odinger representation but only if one substitutes the unbounded 
configuration variable by a bounded polynomial of Weyl operators 
\cite{9}. This latter step is motivated by the full theory where 
it appears in a regularisation procedure and in the limit of vanishing
regulator the original polynomial is recovered. This regulator is 
of point splitting type (technically, it involves the coordinate area
enclosed by a loop that labels a Wilson loop variable) and can be 
arbitrarily small in the full theory but not in LQC and thus stays 
there as a remnant that defines the model. Both in LQC and in full
LQG what survives the limit of taking the regulator away
is the substituting polynomial of Weyl operators 
that one put in. The possible effects of this kind of ambiguity have 
been recently pointed at for LQC in \cite{19}. In LQG they have 
been much debated since their first occurence and there are several 
approaches to fix them. One of them \cite{20,21} is in the context 
of Dirac quantisation of the constraints and uses the representation theory
of the hypersurface deformation algebra. Another one \cite{22} 
exploits the possibility
of gauge fixing before quantisation and applies standard Wilson type 
renormalisation to dynamical systems with a physical Hamiltonian 
(not constrained to vanish) \cite{23}. However, a simple inspection 
of the Mukhanov-Sasaki mass squared term reveals that the 
techniques of \cite{9} for the purely homogeneous contribution 
to the Hamiltonian constraint and of \cite{8} for hybrid LQC
will no longer be sufficient when backreaction is switched on.
This is due to the fact that the adiabatic corrections introduce 
non polynomial functions of {\it both momentum and configuration variables},
in particular, negative powers of both variables occur. 

We solve this second problem by taking an unbiased point of view 
towards quantisation of the homogeneous sector and try to stay within
the standard Schr\"odinger representation as suggested by the Weyl 
quantisation method that enters the SAPT formalism. In the best case 
one should find a dense and invariant domain for the various non
polynomial operators that appear. This is indeed possible for 
the model that also involves Gaussian dust by exploiting the existence 
of a remarkable basis of functions in $L_2(\mathbb{R},dx)$ 
which is smooth and of rapid decrease both at infinity and at the origin 
\cite{24}. For the case of the Mukhanov-Sasaki field we have to content 
ourselves by providing a dense but not invariant domain.\\
\\
This section is organised as follows:\\
\\
In the first subsection we explain in more detail 
why we can consider the homogeneous degrees of freedom 
as the slow sector of 
cosmological perturbation theory.

In the second we display the afore mentioned obstacle to SAPT in the 
QFT context and show how to solve it up to second order in the inhomogeneous
perturbations. A key role is played here by a certain Hilbert-Schmidt 
condition.

In the third we describe the induced non positive, non-polynomial mass squared 
problem and apply both solution techniques sketched above based on either
modifying the classical slow phase space or the number of physical modes in 
the Fock space.

In the fourth subsection we sketch a proposal for how to deal with 
the resulting highly non polynomial operators that occur as a result of 
the canonical transformations and the adiabatic corrections 
due to Moyal product.

\subsection{Homogeneous sector as center of mass degrees of freedom}
\label{s4.1}

In our publications we are considering 4 different models:\\
I. Coupled (an)harmonic oscillators\\
II. Homogeneous isotropic GR coupled to homogeneous mode of the inflaton\\
III. Homogeneous isotropic GR couupled to all modes of the inflaton 
in the presence of 4 deparametrising dust fields\\
IV. GR coupled to an inflaton without deparametrising matter fields\\
\\
Models I. and III. are unconstrained systems and the corresponding 
Hamiltonian is a physical Hamiltonian, in particular, it does not need to 
vanish,
we are interested in the full spectrum and not only its kernel and we must 
not multiply the Hamiltonian with anything. 
Model IV. is a constrained 
system, we are only interested in the kernel of the Hamiltonioan constraint 
and we are allowed to multiply the Hamiltonian with convenient (non-vanishing)
factors if necessary. Model II. can be considered as the truncation of 
both model III and model IV. to the entirely homogeneous degrees of freedom.
Thus it makes sense to consider the full spectrum of the corresponding 
Hamiltonian in model II.  
 
When treating these models with SAPT techniques we notice that what 
makes SAPT work at the technical level is that one multiplies the 
momenta $p$ of the "slow" degrees of freedom $(q,p)$ with a small factor 
$\epsilon$ thus changing the canonical bracket $\{p,q\}=1$ into 
$\{p',q\}=\epsilon$ with $p'=\epsilon$. The factor $\epsilon$ is usually 
physically motivated by a corresponding parameter appearing in the 
Hamiltonian such as a ratio between small and large masses 
$\epsilon^2=m/M$. In the corresponding Moyal products of the corresponding Weyl 
quantisation the parameter $\epsilon$ then organises the perturbative 
expansion of the SAPT scheme. In order that this works, the corresponding
Hamiltonian, when expressed in terms of the primed slow momenta, must 
not contain negative powers of $\epsilon$. Interestingly, when the 
Hamiltonian is a constraint, then we can in fact always achieve that no
negative powers appear by multiplying the constraint by sufficiently large 
powers of $\epsilon$. We will exploit that freedom when treating model 
IV (and, when considered as a truncation of model IV, also model II).    

For model I. the parameter $\epsilon^2$ indeed has the interpretation
of a mass ratio
$m/M$. We notice that the physical intuition here is the equipartition theorem 
which states that in thermodynamical equilibrium at any non vanishing 
temeperature $T$ the phase space average (and therefore the statistical
= time average if the system is ergodic) of both $p^2/M$ and $y^2//m$
is equal. Since $v=p/m,\; u=y/m$ (with $(x,y)$ the fast d.o.f.) 
are the corresponding velocities we find $v=\epsilon u\ll u$. Note that 
this has nothing to do with the frequencies of the oscillators. Even 
if the frequency $\Omega$ of the heavy oscillator is much larger than that 
$\omega$ of the fast  oscilltor we still have $v\ll u$. This is possible 
because $v=b\Omega, u=a \omega$ for amplitudes $b,a$ and we then have 
$b/a\ll \epsilon$.    

For model II. we interpret $\epsilon^2=\kappa/\lambda=:m/M$ where 
$\lambda,\kappa$ 
are the inflaton and gravitational coupling constants. Note that 
$\hbar\kappa$ is the Planck area and $\hbar\lambda$ also has the dimension 
of an area. Note also that $\lambda$ is to be distinguished from the Compton 
wave length $l=\hbar/(\mu c)$ where $\mu$ is the Klein Gordon mass (i.e. 
$2\mu=V^{\prime\prime}(\phi=0)$ for the potential $V$ of the inflaton $\phi$). 
In the SAPT treatment we will assume that $\kappa/\lambda\ll 1$ which 
would be the case if $\hbar\lambda \approx l^2$ which one assumes to be 
of the order of length scale of the standard model. The equipartition theorem 
does not work for this model because in deriving 
\be \label{4.2.1}
<p^2/M>=<y^2/m>,\; <f>:=
\frac{\int\; dq\;dp\;dx\;dy\; e^{-\beta H} \; f}
{\int\; dq\;dp\;dx\;dy\; e^{-\beta H} \; 1}
\ee
one assumes that $H$ is bounded from below so that the integrals converge 
and that in the integrations by parts that one performs to show that 
$<p(\partial H/\partial p)>=\beta^{-1}$
no boundary terms appear. Both conditions are violated in model II because 
the gravitational kinetic energy is negative.
However we use the fact that the Hamiltonian is a constraint. Therefore
up to numerical constants
\be \label{4.2.2}
p^2/(Ma)=M\Lambda a^3+\pi_0^2/(m a^3)+ma^3 V(\phi_0)
\ee
with $p,\pi_0$ the homogeneous momenta of scale factor and inflaton 
respectively. Since $v=\dot{\ln(a)}=-p/(Ma^2),\; \dot{\phi}=u=\pi/(m a^3)$ 
we find 
\be \label{4.2.3}
M v^2=[M\Lambda + m(u^2+V)]
\ee
displaying again $v\ll u$ for very small $\Lambda$ and potential at least.
Accordingly, the zero (homogenous) mode of the inflaton is ``fast'' while 
homogeneous gravity mode is ``slow'' owing to our assumption on 
$\lambda,\kappa$.

Model III. exhibits new challenges due to the inhomogeneous modes of the 
inflaton. Note however, that the inflaton here is a gauge invariant
degree of freedom because the constraints have already been solved 
using reduced phase space methods. In particular, there is no motivation 
from the gauge perspective
to split the inflaton into zero mode (homogenous part) and the rest
as it is the case for the hybrid treatment of model IV. There are no linearised
constraints etc. In order to meet the 
Hilbert Schmidt condition (simultaneous existence of a Fock 
representation supportying the Hamiltonian 
and a representation of the gravitational d.o.f.), after a 
canonical transformation exact up to second order the 
Klein Gordon mass $m^2$ gets modified into 
$\mu^2$ and displays a phase space dependence with respect 
to the homogeneous gravitational d.o.f. but {\it not} the zero mode of the 
inflaton, it does not depend on the inflaton d.o.f. at all. 
That mass dependence is such that $p$ only appears in the combination 
$p^2 \kappa^2=\epsilon^4 \lambda^2 p^2=\epsilon^2 \lambda^2 (p')^2$. i.e.
only positive powers of $\epsilon$ appear in $\mu^2$ when expressed in terms 
of $p'$. Since the Hamiltonian is still not bounded from below, the 
equipartition theorem can again not be used to argue that the 
$(p,a)$ are slow compared to all inflaton modes. However, the physical 
Hamiltonian $H$ results from a constraint of the form $C=P+H=0$ where 
$P$ is the energy density of the dust. As the dust behaves closely to a 
field of test observers with zero energy density, P is very small and thus 
$H$ is close to zero. Thus we can argue and apply the SAPT scheme 
as for model II if we assume that all inflaton modes are at least as 
fast as its homogeneous ones. See below for a justification.

Finally in model IV. we treat both the gravitational tensor modes (graviton) 
and 
the gauge invariant extension of the inflaton (Mukhanov-Sasaki (MS)
field $\nu$) 
using the hybrid 
scheme (the vector modes are not gauge invariant to first odrder and 
are dropped). Due to the requirement of linearised gauge invariance, the 
mass squared term $\mu^2$ of $\nu$ as well as the mass squared term $\rho^2$ 
of the  graviton acquire a 
phase space dependence wrt both the homogneous inflaton as well as homogenous 
gravitational dof. Additional phase space dependence of $\mu^2,\rho^2$ 
with respect to these d.o.f. is generated  by a HS condition guaranteeing 
canonical transformation as for model III. Due to the additional dependence of
$\mu^2,\rho^2$ on the homogeneous inflaton d.o.f. $(\pi_0,\phi_0)$ 
it is necessary to show that 
these also ``slow'' as compared to the inhomogeneous d.o.f.
$(\nu,\pi_\nu)$ and $(h^{ab},\pi_{ab})$ if we want to 
use the SAPT scheme.
The graviton piece of the Hamiltonian reads (up to powers of $a$)
\be \label{4.2.4}
\kappa \pi^{ab} \pi_{ab}+h^{ab}(-\Delta/\kappa+\rho^2) h_{ab}
\ee
To make it look like the MS piece
\be \label{4.2.5}
\lambda \pi_\nu^2+\nu(-\Delta/\lambda+\mu^2)\nu
\ee
we perform the canonical transformation 
$\tilde{\pi}^{ab}=\epsilon \pi^{ab},\;
\tilde{h}_{ab}=\epsilon^{-1} h_{ab}$ so that (\ref{4.2.4})  becomes 
\be \label{4.2.6}
\lambda\tilde{\pi}^{ab} \tilde{\pi}_{ab}
+\tilde{h}^{ab}(-\Delta/\lambda+\tilde{\rho}^2) \tilde{h}_{ab}
\ee
where $\tilde{\rho}^2=\epsilon^2\rho^2$. 
An explicit check reveals that both the MS mass $\mu^2$ as well 
as $\tilde{\rho}^2$ receive only non negative powers of $\epsilon$ when 
expressed in terms of 
\be \label{4.2.7}
\tilde{p}=\epsilon^2 p=\epsilon p',\; \tilde{\pi_0}=\epsilon \pi_0
\ee
Note that while in model II we used $p',\pi'_0=\pi_0$ instead of 
$\tilde{p},\tilde{\pi}$ we still have $\tilde{p}/\tilde{\pi}=p'/\pi'$. 
This corresponds to the fact that when multiplying model II viewed as 
the homogenous truncation of model IV by $\epsilon^2$
we could have as well worked with $\tilde{p},\tilde{\pi}$. 

The only problem is that the homogeneous piece of the Hamiltonian constraint 
of model IV cannot be written in terms of $\tilde{p},\tilde{\pi}$ without 
picking up negative powers of $\epsilon$. But this can be repaired, as 
remarked above and following the observation just made by multiplying 
the {\it entire} constraint including the inhomogeneous MS and graviton 
pieces by $\epsilon^2$. We now perform again a canonical transformation 
$\hat{\pi}_\nu=\epsilon \pi_\nu,\; \hat{\nu}=nu/\epsilon$ and similar 
for the graviton variables to write the constraint in terms of 
\be \label{4.2.8}
\lambda \hat{\pi}_\nu^2+\hat{\nu}\epsilon^4(-\Delta/\lambda+\mu^2)\hat{\nu}
\ee
and similar for the graviton piece. Defining $\omega^2=-\Delta/\lambda+\mu^2$
we now define the Fock representation by the annihilation operator
\be \label{4.2.9}
\hat{b}=[\epsilon\sqrt{\omega}\hat{\nu}-i(\epsilon\sqrt{\omega})^{-1}
\hat{\pi}_\nu)/\sqrt{2}
\ee
Remarkably $\hat{b}=b$ is exactly the same annihilation operator as 
one would have defined before that last canonical transformation (i.e.
resubstituting $\nu,\pi_\nu$), i.e. the Fock representations are 
{\it identical}. 
When normal ordered one finds that the MS Hamiltonian becomes 
\be \label{4.2.10}
\epsilon^2\; \int\; d^3x\; b^\dagger\omega^2 b
\ee
ie. the spectrum of the inhomogeneous part of the constraint 
gets simply rescaled by $\epsilon^2$ and one can use all the 
results of the original Fock representation. This is in fact neat as one 
would expect that the homogeneous modes alone as correspoonding to 
model II multiplied by $\epsilon^2$ are little disturbed by the 
inhomogeneous ones. 

In summary, all one has to do is a simpe rescaling 
$p\to \epsilon^2 p=\tilde{p},\;
\pi \to \epsilon \pi-\tilde{\pi}$ in order to use the SAPT scheme in model IV 
and 
this is consistent with the treatment of model II.

In fact, we can also treat model III consistently this way by simply 
multiplying (and, to get the correct spectrum, afterwards dividing) 
by $\epsilon^2$ thus working in all
models II., III. and IV. consistently with $\tilde{p},\tilde{\pi}$.
\\
\\
What is missing is the justification for why the homogeneous scalar field mode 
should be 
slower than the inhomogeneous modes in model IV and for completeness also 
in model III although there it would be sufficient that all inflaton modes 
are equally fast. It turns out that the answer to the question lies 
in the definition of the modes. By homogeneous mode we losely speaking 
mean a component of the field which does not depend on position. But 
this becomes unambiguous only when relating it to the full field.
Given the torus of volume $L^3$ we consider the mode system 
$e_I,\; I\in \mathbb{Z}^3,\; e_I(x)=\exp(i k_L I\cdot x),\;
k_L=2\pi/L$ which enjoy 
\be \label{4.2.11}
<e_I, e_J>= L^3 \delta_{I,J},\;\frac{1}{L^3}\sum_n\; e_I \;<e_I,.>=1_{L_2},\;
<f,g>=\int_{[0,L]^3}\;d^3x\; f^\ast(x)\;g(x)
\ee
periodic boundary conditions understood. We define the homogenous modes 
of say a scalar field $\phi,\pi$ by
\be \label{4.2.12}
\overline{\phi}:=\frac{1}{L^3}\int \; d^3x\; \phi(x)=<e_0,\phi>/L^3,\;\;   
\overline{\pi}:=\frac{1}{L^3}\int \; d^3x\; \pi(x)=<e_0,\pi>/L^3\;\;   
\ee
which are easily checked to have canonical brackets
$\{\overline{\pi},\overline{\phi}\}=\frac{1}{L^3}$ if 
$\{\pi(x),\phi(y)\}=\delta(x,y)$. We see already that the local point 
modes $\phi(x),\pi(x)$ are ``infinitely faster'' than the homogenous modes
because $\delta(x,x)=\infty$. To understand this in more detail we note that
SAPT was developed for quantum mechanical systems with a finite number 
of degrees of freedom. Acoordingly we regularise the Hamiltonian 
using a finite resolution cut-off $\delta_N=L/N$ and a finite number of
position localised degrees of freedom $\phi^N_I=\int\; d^3x \;
\chi_{N,I}(x)\; \phi(x)/delta_N^3$ and similar for $\pi^N_n$ where 
$\chi_{N,I},\;
I\in \mathbb{N}_N^3,\;\mathbb{N}_N=\{0,1,2,..,N-1\}$ is a partition 
of the torus into disjoint cubes of volume $\delta_N^3$. We check that 
$\{\pi^N_I,\phi^N_J\}=\delta_{I,J}/\delta_N^3$ so that $\overline{\phi}$
is slower than any of the $\phi^N_I$ by a factor of $N^3$. In fact we note 
the identity $\overline{\phi}=\frac{1}{N^3}
\sum_{I\in \mathbb{N}_N^3}\; \phi^N_I$. We see that $\overline{\phi}$ is the 
{\it centre of mass coordinate} of a system of $N^3$ coordinates $\phi^N_I$
of equal mass. Thus at finite resolution we have an abstract gas of 
interacting particles with ``position'' coordinate $\phi^N_I$ and it is 
well known from classical mechanics
that the centre of mass coordinate acqiures the total mass of all 
particles as its effective mass, thus making it much heavier than the 
individual particles. 
  
To see this in more detail, due to the above identity not all of the 
$\phi^N_I$ are 
independent. In case that the KG mass is small, the potential 
term in the Hamiltonian is well approximated by just the discretised Laplacian
\be \label{4.2.13}
-\frac{\delta_N^3}{\delta_N^2}\sum_I\;\phi_I\sum_{a=1}^3(\phi^N_{I+\delta_a}+
\phi^N_{I-\delta_a}-2\phi^N_I)
=\delta_N\sum_I\;\sum_a (\phi^N_{I+\delta_a}-\phi^N_I)^2
\ee
which only depends on the relative coordinates. Here 
$\delta_a$ is the standard basis of $\mathbb{R}^3$. It is thus motivated 
to choose new canonical variables 
$q_I:=\phi^N_I-\phi^N_0,\; I\not=0$ and $\bar{q}:=\overline{\phi}$. 
Then with $A=N^3$ and
$\phi^N_0=\overline{q}-\sum_{I\not=0} q_n/A$ and the symplectic 
potential is
\be \label{4.2.14}
\frac{\theta}{\delta_N^3}=
\sum_{I\not=0} [\pi^N_I-\overline{\pi}]\; dq_n+[A\overline{\pi}]\;d\overline{q}
\ee
Thus the momentum conjugate to $\overline{q}$ is the total momentum
$P=\sum_n \pi_I^N=\overline{\pi}$ while 
$p_I=\pi^N_I-\overline{\pi}$ is conjugate to $q_I$. It follows
$\sum_{I\not=0}p_I=\bar{\pi}-\pi^N_0$ and 
twice the kinetic term in the Hamiltonian at finite cut-off is given by 
\be \label{4.2.15}
\int\; d^3x \; \pi(x)^2\approx \sum_I (\pi^N_I)^2\;\delta_N^3
=\delta_N^3\;[\frac{P^2}{A}
+\sum_{n\not=0} p_n^2+(\sum_{n\not=0}p_n)^2]
\ee
which displays the large total mass $A$ of the centre of mass momentum $P$.
A little bit of further analysis shows that the quadratic form of the 
$p_I$ can be diagonalised by a a furher orthogonal transformation (which can 
be extended to a canonical one) displaying  $A-2$ modes of unit mass and 
one mode of the ``reduced mass'' $1/A$. Thus the centre of mass mode is indeed
by far the heaviest d.o.f. especially in the limit $A\to \infty$.

One may object that the modes $\phi^N_I$ have nothing to do with the 
momentum modes $\hat{\phi}_I=<e_I,\phi>$ except for the zero mode 
$\hat{\phi}_0=A \bar{q}$. However, this is not true because 
the $\phi^N_I$ are approximants of $\phi(x=\delta_N I)$, i.e. the field
discretised on a lattice of the compact space given by the torus of volume
$L^3$. As is well known, in this situation, the Fourier transformation 
can also be restricted to the modes $k_I=k_L I=\frac{2\pi}{L} I,\;
I\in \mathbb{N}_N^3$. We can in fact see this explicitly by writing
\be \label{4.2.16}
\phi^N_I=\frac{1}{L^3\delta_N^3}\sum_{I'\in \mathbb{Z}^3}
\;
<e_{I'},\chi_{N,I}>\; \hat{\phi}_{I'}
\ee
It is easy to see that the Fourier coefficients satisfy
$<e_{I'},\chi_{N,I}>=<e_{I'+l N},\chi_{N,I}>$ for any $l\in \mathbb{Z}^3$
so that (take $N$ even when taking $N\to \infty$ w.l.g.)
\be \label{4.2.17}
\phi^N_n=\frac{1}{L^3\delta_N^3}\sum_{n'\in \mathbb{Z}_{N/2}^3}
\;
<e_{n'},\chi_{N,n}>\sum_{l\in \mathbb{Z}^3}\; \hat{\phi}_{n'+ l N}
\ee
where $\mathbb{Z}_{N/2}=\{-N/2,-N/2+1,..,N/2 -1\}$.
Thus, corresponding to the position space cutoff at $\delta_N$ we perform 
a momentum space cut-off at $N k_L=2\pi/\delta_N$ and set 
$\hat{\phi}_{I'+l N}=0$ for $l\not=0$. In the limit $N,n\to \infty,\;
x=I\delta_N$ fixed, we recover the continuum relation 
$\phi(x)=\frac{1}{L^3}\sum_{I\in \mathbb{Z}^3} e_I(x)\; \hat{\phi}_I$.
At finite $N$ the (as one can show non-singular) 
matrix with matrix elements $<e_{I'},\chi_{N,I}>/\delta_N^3$ 
transforms between
the position and momentum space discretisations. 

In summary we have shown that the zero mode can be considered as 
a centre of mass mode with respect to certain linear combinations 
of discretised position modes which in turn are linear combinations 
of discretised momentum modes. The relation
\be \label{4.2.18}
\sum_{n\in \mathbb{N}_N^3} \phi^N_n/N^3=\overline{\phi}=<e_0,\phi>/L^3\
\ee
remains exaxt also at finite $N$.
Accordinly, treating the homogeneous mode as the by far most massive one
is physically justifyable from this point of view. That instead of an 
arbitrarily large relative scale 
$1/N^3\to 0$ we just used the finite one $\epsilon^2=\kappa/\lambda$
is motivated by the specific combinations of the homogeneous momenta that 
appear in the Hamiltonian.

\subsection{Quantum Cosmological SAPT}
\label{s4.2}

In order to understand the source of the problem originating 
from the infinite number of degrees of freedom, it is illustrative 
to consider a scalar field $\phi$ with conjugate momentum $\pi$ 
coupled to Gaussian dust and General Relativity, that is model III
described above. When gauge 
fixing the spacetime diffeomorphism gauge freedom the gravitational 
and scalar contribution to the Hamiltonian constraint combine to 
a physical Hamiltonian when integrated over the spatial slice
\cite{5}. All gravitational and scalar degrees of freedom now 
become observable fields. To simplify the model we switch off the 
inhomogeneous gravitational modes wrt a flat FRW background 
by hand. The Hamiltonian then takes
the form 
\be \label{4.1.1}
h=h_h(a,p,\phi_0,\pi_0)+h_i(\phi',\pi';a),\;\;
h_i(\phi,\pi;a):=\frac{1}{2}\int_{\sigma}\;
d^3x\; [\frac{\pi^2}{a^3}+a^3\;\phi(\frac{-\Delta}{a^2}+m^2)\phi]
\ee
with purely homogeneous contribution $h_h$ whose precise form 
is not important for the purpose of this subsection and a 
contribution $h_i$ which depends on the inhomogeneous scalar field modes 
$(\phi',\pi')$ as well as the scale factor $a$.
Here $p$ denotes the momentum conjugate to $a$, $(\phi_0,\pi_0)$ are 
the homogeneous scalar field modes and $\sigma\cong T^3$ 
is diffeomorphic to a 3-Torus of unit coordinate volume and Laplacian 
$\Delta$. Note that the split (\ref{4.1.1}) is not with respect 
to the adiabatic parameter.

When quantising $h$ using the formalism of SAPT we are asked to work
on the Hilbert space ${\cal H}={\cal H}_f\otimes {\cal H}_s$. As far
as the slow sector is concerned we will adopt a usual Schr\"odinger 
representation in accordance with the SAPT formalism. As $h_i$ 
is quadratic in $(\phi',\pi')$ a Fock representation suggests itself,
{\it but which one}? After all, the ``background variable'' $a$ is not a 
fixed function of time but rather a dynamical variable which is
in addution quantised, hence displays quantum fluctuations. The SAPT
formalism enables us to consider $a$ as a real parameter when quantising 
$H_i$ with respect to the fast variables $(\phi',\pi')$, hence 
it is natural to consider the Fock space ${\cal H}_f(a)$ with vacuum 
$\Omega_a$ and annihilation operator valued distribution symbol
\be \label{4.1.2}
b_a(x):=\frac{1}{\sqrt{2}}[\sqrt{\omega_a}\phi'-
i\sqrt{\omega_a}^{-1}\pi'](x),\;
\omega_a^2=-\Delta a^2+m^2 a^6
\ee
where $x\in \sigma$ because then the {\it $a$ dependent normal ordering}   
of $H_i$ reads
\be \label{4.1.3}
h_i(a)=\frac{1}{a^3}\int_\sigma\; d^3x\; b_a^\ast\;\omega_a b_a
\ee
Then the excitations of $\Omega_a$ are obtained by applying the 
smeared creation operators 
\be \label{4.1.4}
b_a(f)^\ast=<A_a,f>=\int_\sigma\; d^3x\; A_a^\ast\; f
\ee
where $f$ is smooth. 

Two immediate questions pose themselves:\\
1.\\ 
Are the corresponding Fock representations $(\rho_a,{\cal H}_f(a))$ of the 
$(\phi',\pi')$ all unitarily equivalent to a single 
one $(\rho_f,{\cal H}_f)$? This is one of the innocent looking
assumptions of SAPT which is automatically satisfied in the quantum 
mechanical context. \\
2.\\
Assuming that this is the case, 
let $B(f)=W(b(f)),\;B^\ast=W(b(f)^\ast)$ be the Weyl quantisations of the 
symbols $b(f),b(f)^\ast$. Is the complete 
algebra of the operators $a,p,\phi_0,\pi_0,B(f),B(f)^\ast$
still well defined on ${\cal H}_f\otimes {\cal H}_s)$?\\
\\
It turns out that both questions are tightly related and surprisingly, 
the answer to both is negative. The underlying effect has been first observed 
in \cite{8} in a related context. To see the origin of the problem, we 
note that a necessary condition for an affirmative answer to the first 
question is that the Fock vacuum $\Omega_{a_2}$ can be written as 
an excited state in ${\cal H}_{a_1}$ for all $a_1,a_2$? In fact, this 
condition is also sufficient beccause polynomials of the $b_{a_1}(f)^\ast$
can be written as polynomilas of the $b_{a_2}(f),b_{a_2}(f')$. Accordingly
we make the Ansatz
\be \label{4.1.5}
\Omega_{a_2}=\sum_N\; z_N\; e_{a_1}(N)\;\;
e_{a_1}=\prod_I\; \frac{[b_{a_1}(f_I)^\ast]^{N_I}}{\sqrt{N_I !}}
\Omega_{a_1}
\ee
where $I$ labels an orthornormal basis of 
inhomogeneous mode functions 
on $\sigma$
\be \label{4.1.7}
<f_I,f_J>=\int_{\sigma}\; d^3x\; \overline{f_i(x)}\;f_J(x)
\ee
and $N$ denotes the collection of the occupation numbers $N_I$. The 
sum is over all $N$ with only finitely many $N_I$ different from zero.
Then we impose $b_{a_2}(f_I)\Omega_{a_2}=0$ by using the 
Bol'ubov decomposition
\ba \label{4.1.8}
b_{a_2} &=& \kappa_+ b_{a_1} + \kappa_- b_{a_1}^\ast
\nonumber\\
\kappa_\pm=
\frac{1}{2}[
\sqrt{\frac{\omega_{a_1}}{\omega_{a_2}}}
\pm \sqrt{\frac{\omega_{a_2}}{\omega_{a_1}}}
\ea
As in our case all operators $\omega_a$ are mutually commuting we can 
pick the $f_I$ to be eigenfunctions of $-\Delta$ with eigenvalues 
$k_I^2$. Specifically on the torus the label set is $\mathbb{Z}^3-\{0\}$
and $k_I^2 = (2\pi)^2||I||^2/L^2$ where $L^3$ is the volume of the 
torus. Let $\kappa_\pm(I)$ be the associated eigenvalues of 
$\kappa_\pm$ and let 
$\delta_I$ be the the occupation number configuration $(\delta_I)_J=
\delta_{IJ}$ then 
\ba \label{4.1.9}
0&=& b_{a_2}(f_I)\Omega_{a_2}
\nonumber\\
&=&\sum_N\; z_N\; [
\sqrt{N_I}\; <f_I,\kappa_+ f_I> \; e_{N-\delta_I}
+\sqrt{N_I+1}\; <f_I,\kappa_- f_I> \; e_{N+\delta_I}]
\nonumber\\
&=&\sum_N\; [
z_{N+\delta_I}\;\kappa_+(I)\sqrt{N_I+1}+
z_{N-\delta_I}\;\kappa_-(I)\sqrt{N_I}
]\;e_N
\ea
Since this hold for all $I$ independently, the coefficients must be of 
infinite product type
\be \label{4.1.10}
z_N=\prod_I z^I_{N_I}
\ee
which transforms (\ref{4.1.9}) into the recursion
\be \label{4.1.11}
z^I_{N+1}=-\sqrt{\frac{N}{N+1}} \sigma_I\; z^I_{N-1},\; 
\sigma_I:=\frac{\kappa_-(I)}{\kappa_+(I)}
\ee
where the r.h.s. vanishes for $N=0$. It follows $z^I_N=0$ for $N$ odd 
while the solution of (\ref{4.1.11}) for $N$ is given by
\be \label{4.1.12}
z^I_{2N}=-\sqrt{\frac{2N-1}{2N}}\sigma_I\; z^I_{2(N-1)}
=(-\sigma_I)^N \sqrt{\frac{(2N)!}{4^N (N!)^2}} z^I_0
\ee
where $z^I_0$ remains undetermined.
It follows using manipulations well known from statistical physics
\ba \label{4.1.13}
1&=& ||\Omega_{a_2}||^2_{{\cal H}_{a_1}}
=\sum_N |z_N|^2=\sum_N [\prod_I |z^I_{N_I}|^2]=
\prod_I |z^I_0|^2\; |[\sum_{N=0}^\infty |z^I_N|^2]
\nonumber\\
&=& [\prod_I |z^I_0|^2]
\prod_I [\sum_{N=0}^\infty \sigma_I^{2N}\; \frac{(2N)!}{4^N (N!)^2}]
\nonumber\\
&\ge & [\prod_I |z^I_0|^2]
\prod_I [\sum_{N=0}^\infty (\frac{\sigma_I^2}{2})^N]
= [\prod_I |z^I_0|^2]
\prod_I (1-\sigma_I^2/2)^{-1}]
\nonumber\\
&\le & [\prod_I |z^I_0|^2]
\prod_I [\sum_{N=0}^\infty \sigma_I^{2N}]
= [\prod_I |z^I_0|^2]
\prod_I (1-\sigma_I^2)^{-1}]
\nonumber\\
\ea
where the basic estimates $1\ge (2N)!/(4^N (N!)^2)\ge 2^{-N}$ were used.
Note that $\sigma(I)<1$ so that the infinite product in the last step
coming from the geometric series is meaningful.  
Thus a necessary condition for 
convergence of (\ref{4.1.13} is that the two infinite 
products converge independently to a finite, non zero value since 
$\prod_I z^I_0$ is a common prefactor in all $z_N$ thus it must be 
convergent to some value $Z$ by itself as otherwise all $z_N$ would be 
meaningless. 
By taking the logarithm, the convergence of the second 
infinite product is equivalent to the convergence of the series
\be \label{4.1.14}
\sum_I \ln(1-\sigma_I^2/2) 
\ee
known as the {\it Hilbert-Schmidt} condition.
We have with the abbreviation $\omega_j=\omega_{a_j}(I)$
\ba \label{4.1.15}
\sigma_I^2&=&\frac{\kappa_-(I)^2}{\kappa_+(I)^2}
=\frac{\frac{\omega_1}{\omega_2}+\frac{\omega_2}{\omega_1}-2}
{\frac{\omega_1}{\omega_2}+\frac{\omega_2}{\omega_1}+2}
\nonumber\\
&=&
[\frac{\omega_1-\omega_2}{\omega_1+\omega_2}]^2
=[(\frac{\omega_1^2-\omega_2^2)^2}{(\omega_1+\omega_2)^4}
\nonumber\\
&=& 
=[(\frac{(a_1^4-a_2^4)k_I^2+(a_1^6-a_2^6)m^2)^2}{(\omega_1+\omega_2)^4}
\ea
For large $k_I$ we have $\omega_j\propto k_I a_j^2$ thus the fraction in 
\ref{4.1.15}) approaches a constant and the series in (\ref{4.1.14}) 
diverges for any $a_1 \not= a_2$. On the other hand, note that if the 
coefficient in front of $k_I^2$ in $\omega_a^2$ would not depend on 
$a$ then $\sigma_I^2$ and thus $\ln(1-\sigma_I^2/2))$ would decay as 
$1/k_I^4$ and thus the series 
\be \label{4.1.16}
\sum_I \ln(1-\sigma_I^2) 
\ee
would converge which would be
sufficent for the convergence of (\ref{4.1.13}) to a non zero value.\\
\\ 
This answers the first question posed above. To see that the 
second question is equivalent to the first we note that given that 
$a$ is represented as a self-adjoint operator $A$ on the full Hilbert
space $\cal H$, by the spectral theorem we may display the Hilbert space
as a direct integral or Hilbert bundle subordinate to  $A$
\be \label{4.1.17}
{\cal H} \cong \int_{\mathbb{R}^+_0}^\oplus \; d\mu(a) \; {\cal H}_a
\ee
where we identify the fibre spaces ${\cal H}_a$ as the Fock spaces 
${\cal H}_f(a)$ considered above and $\mu$ is the spectral probability
measure 
on the spectrum of $A$ which is $\sigma(A)=\mathbb{R}^+_0$.
As a consequence of the spectral theorem
the ${\cal H}_a$ (of equal dimension) can be chosen identical \cite{25a}
which we already know to be not the case but is instructive to pretend 
to not know this. Vectors in the Hilbert bundle are given by measurable fibre 
Hilbert space valued functions $\psi: \sigma(A)\mapsto {\cal H},\; a\mapsto
\psi(a)$ over the base manifold $\sigma(A)$ and are 
equipped with the inner product
\be \label{4.1.18}  
<\psi,\psi'>=\int\; d\mu(a)\; <\psi(a),\psi'(a)>_{{\cal H}_a}
\ee
By the spectral theorem, $A$ acts by multiplication by $a$ in the 
fibre ${\cal H}_a$ and accordinly the operator $H_i$ acts fibre wise
as well by the symbol $h_i(a)$ in (\ref{4.1.3}). The question is now 
how the operator $P$ representing $p$ acts on the direct integral Hilbert 
space. As the spectrum of $A$ is of absoltely continuous type, it acts 
as $P\psi=\psi'$ where $\psi'(a)=[i \hbar d/da+f(a)]\psi(a)$ where $f(a)$ 
is related to the divergence of the measure $\mu$ and turns $P$ into a 
symmetric operator (in fact, in order to obtain a self-adjoint operator one 
should pass to the real valued triad variable $e$ and work with its conjugate 
momentum but the conclusion derived below is not affected by these 
sutleties). By contrast the operators $B,B^ast$ act fibre wise by 
the corresponding symbols.

It follows
\be \label{4.1.19}
[P,B]=i\kappa\cdot B^\ast,\;[P,B]=i\hbar K'\cdot B,\;
\kappa'(a)= \frac{d}{da} \ln(\sqrt{\omega_a})
\ee
where $K'$ is the Weyl quantisation of the symbol $\kappa'$
and accordingly
\be \label{4.1.20}
B\; P\; \Omega=[B\; P]\; \Omega=-i\hbar K' \cdot B^\ast \Omega
\ee
which is solved by 
\be \label{4.1.21}
P\Omega=-i\frac{\hbar}{2}\int_\sigma d^3x B(x)^\ast (K'\cdot B)^\ast(x)
\Omega
\ee
The vector (\ref{4.1.20}) has the norm 
\be \label{4.1.22}
||P\Omega||^2
=\int_{\sigma(A)}\; d\mu(a)\; Tr_{L_2(T^3)}((\kappa'(a))^\dagger \kappa'(a))
\ee
which is finite only if the {\it Hilbert-Schmidt} norm
\be \label{4.1.23}
Tr_{L_2(T^3)}((\kappa'(a))^\dagger \kappa'(a))=
\sum_I |<f_I,\kappa'(a) f_I>|^2 =\frac{1}{16} 
=\sum_I [\frac{\frac{d}{da} \omega_I(a)^2}{\omega_I(a)^2}]^2
\ee
is finite $\mu$ a.e.
We easily see that (\ref{4.1.23}) is the infinitesimal 
version of (\ref{4.1.14}) if we divide it by $(a_1-a_2)^2$ and 
take the limit $a_2\to a_1$.\\
\\
Thus it is not possible to apply SAPT to the given Hamiltonian directly.
One may think that by a different choice of Fock representations 
one maybe able to satisfy the Hilbert-Schmidt condition. However,
if one wants to keep the correspondingly normal ordered Hamiltonian 
at least densely defined on Fock states then this again leads to 
an obstruction as one can show with more work. 

To understand the origin of this 
obstruction, note that we can satisfy the Hilbert-Schmidt
condition if we rescale the inhomogeneous degrees of freedom 
as
\be \label{4.1.24}
\tilde{\phi}'=a \phi',\;   
\tilde{\pi}'=\frac{\pi'}{a}
\ee
which still have canonical brackets and in terms of which we have 
\be \label{4.1.25}
h_i=\frac{1}{2a}\; \int_\sigma\; d^3x\; [(\tilde{\pi}')^2+
\tilde{\phi}'\tilde{\omega}_a^a \tilde{\phi}'],\;\;
\tilde{\omega}_a^2=-\Delta+m^2 a^2
\ee
so that the coefficient of $-\Delta$ in $\tilde{\omega}_a^2$ is 
independent of $a$. However, (\ref{4.1.24}) is not a canonical transformation 
on the full phase space so that it is no longer the case that 
$p,\tilde{\phi}',\tilde{\pi}'$ have vanishing Poisson brackets.
Consequently, $p$ cannot simply act on the $a$ dependence of a wave 
function. One can of course complete the transformation (\ref{4.1.24})
{\it exactly} by adding a corresponding contact term in the symplectic potential
\be \label{4.26}
\Theta=
p da+\pi_0 d\phi_0+\int\; dx\; \pi' d\phi'¨= 
(p-\frac{1}{a}\int dx \pi' \phi') da+\pi_0 d\phi_0
+\int\; dx\; \tilde{\pi}' d\tilde{\phi}' 
\ee
displaying
\be \label{4.1.27}
\tilde{p}=p-\frac{1}{a}\int dx \pi' \phi'),\; \tilde{a}=a,\;
\tilde{\phi}_0=\phi_0,\;   
\tilde{\pi}_0=\pi_0
\ee
as the completion of that transformation. Unfortunately, now we have to 
write $h_h$ in terms of $\tilde{p}$ thereby introducing first and second 
powers of the (normal ordered) operator 
\be \label{4.1.28} 
\int dx \pi' \phi'=
\int dx \tilde{\pi}' \tilde{\phi}'=
\frac{i}{2}\int dx[\tilde{b}^2-(\tilde{b}^\ast)^2-2\tilde{b}^\ast \tilde{b}]
\ee
Here $\tilde{b}$ is the annihilator obtained from (\ref{4.1.2}) by 
substituting all ingredients by those with a tilde. The operator 
(\ref{4.1.29}) is obviously ill defined on the corresponding Fock space.

However, the discussion suggests to consider more general canonical 
transformations in order to avoid the desastrous terms such as (\ref{4.1.28}).
To restrict the class of such transformations we follow the logic of 
\cite{8}: We remember that 
at present we are interested in perturbation theory with respect to 
the inhomogeneities up to second order in $\phi',\pi'$ which themselves are 
considered to be of first order. This suggests to restrict to transformations 
which are linear in $\phi',\pi'$ such as (\ref{4.1.24}) 
since this keeps the second order 
nature of $h_i$ and higher order transformations would anyway 
not be visible at the second order precision of $h_i$. The corresponding
contact terms for the homogeneous degrees of freedom will then be 
already of second order in leading order as in (\ref{4.1.27}) and we can 
restrict the precision of the canonical transformation to second order.

Correspondingly we consider transformations of the form (local in $x$)
\be \label{4.1.29}
\phi'=r\cdot \tilde{\phi}' +s\cdot \tilde{\pi}',\;
\pi'=t\cdot \tilde{\phi}' +u\cdot \tilde{\pi}',\;
\ee
where $r,s,t,u$ in principle can depend on all homogeneous degrees of freedom.
Also all functions may involve a non-trivial (translation invariant - if we 
want to keep translation covariance) integral kernel (which is 
why we use the $\cdot$ notation) and to satisfy 
the Hilbert Schmidt condition it will be suffcient to restrict it to
be constructed from $\Delta$ so that they mutually commute and are symmetric 
on $L_2(T^3)$. Of course, $r,s,t,u$ are restricted to be real valued
since all variables are. In order that (\ref{4.1.29}) be canonical 
taking only the Poisson brackets between $\phi',\pi'$ into account 
we must have 
\be \label{4.1.30}
-t\cdot s+r\cdot u=1_{L_t(T^3)} \;\;\Rightarrow\;\;
u\cdot r-s\cdot t=1_{L_t(T^3)}
\ee
where the symmetry of the kernels was exploited and that 
$[r\cdot,s\cdot]=[t\cdot,u\cdot]=0$ due to mutual commutativity and 

As before one now plugs (\ref{4.1.29}) into the symplectic 
potential and computes 
the contact terms up to total differentials to second order in the 
perturbations. One then expresses the Hamiltonian in terms of the new 
variables, expands it to second order in the perturbations and 
determines the functions $r,s,t,u$ such that the terms not densely 
defined on the Fock space cancel each other and such that the 
Fock spaces determined for different values of the homogeneous variables
are all identical which will be the case if and only if the coefficient
of $-\Delta$ in the corresponding frequency squared operator is independent
of the phase space variables.   

We will derive the completion of the canonical tranformation (\ref{4.1.29})
abbreviating the homogeneous canonical pairs by $q^j,p_j,\; j=1,2$, using 
that convoluted kernels are also symmetric 
and dropping total differentials as well as terms of fourth order in the 
perturbations 
\ba \label{4.1.31}
\Theta &= & 
p_j dq^j + \pi'\cdot d\phi' 
=p_j dq^j + [t\cdot\tilde{\phi}'+u\cdot\tilde{\pi}']\cdot 
d[r\cdot\tilde{\phi}'+s\cdot \tilde{\pi}'] 
\nonumber\\
&=& p_j dq^j +\tilde{\pi}'\cdot(u\cdot r- t\cdot s)\cdot d\tilde{\phi}'
\nonumber\\
&& -\frac{1}{2}[\tilde{\phi}'\cdot d(t\cdot r)\cdot \tilde{\phi}'
+\tilde{\pi}'\cdot d(u\cdot s)\cdot \tilde{\pi}'+
+2 \tilde{\phi}'\cdot d(t\cdot s)\cdot \tilde{\pi}']
\nonumber\\
&& +[\tilde{\phi}'\cdot(t\cdot dr)\cdot \tilde{\phi}'+
\tilde{\phi}'\cdot(t\cdot ds+u\cdot dr)\cdot \tilde{\pi}'
+\tilde{\pi}'\cdot (u\cdot ds)\cdot \tilde{\pi}']
\nonumber\\
&=& p_j dq^j +\tilde{\pi}'\cdot d\tilde{\phi}'
\nonumber\\
&& -\frac{1}{2}[\tilde{\phi}'\cdot (t\cdot r)_{,q^j} \cdot \tilde{\phi}'
+\tilde{\pi}'\cdot (u\cdot s)_{q^j}\cdot \tilde{\pi}'+
+2 \tilde{\phi}'\cdot (t\cdot s)_{,q^j}\cdot \tilde{\pi}']\; dq^j
\nonumber\\
&& -\frac{1}{2}[\tilde{\phi}'\cdot (t\cdot r)_{,p^j} \cdot \tilde{\phi}'
+\tilde{\pi}'\cdot (u\cdot s)_{p_j}\cdot \tilde{\pi}'+
+2 \tilde{\phi}'\cdot (t\cdot s)_{,p_j}\cdot \tilde{\pi}']\; dp^j
\nonumber\\
&& +[\tilde{\phi}'\cdot(t\cdot r_{,q^j})\cdot \tilde{\phi}'+
\tilde{\phi}'\cdot(t\cdot ds+u\cdot r_{,q^j})\cdot \tilde{\pi}'
+\tilde{\pi}'\cdot (u\cdot s_{,q^j})\cdot \tilde{\pi}']\;dq^j
\nonumber\\
&& +[\tilde{\phi}'\cdot(t\cdot r_{,p_j})\cdot \tilde{\phi}'+
\tilde{\phi}'\cdot(t\cdot ds+u\cdot r_{,p_j})\cdot \tilde{\pi}'
+\tilde{\pi}'\cdot (u\cdot s_{,p_j})\cdot \tilde{\pi}']\;dp_j
\nonumber\\
&=& \tilde{\pi}'\cdot d\tilde{\phi}'
\nonumber\\
&& +[p_j
-\frac{1}{2}[\tilde{\phi}'\cdot (t\cdot r)_{,q^j} \cdot \tilde{\phi}'
+\tilde{\pi}'\cdot (u\cdot s)_{q^j}\cdot \tilde{\pi}'+
+2 \tilde{\phi}'\cdot (t\cdot s)_{,q^j}\cdot \tilde{\pi}']
\nonumber\\
&&+
[\tilde{\phi}'\cdot(t\cdot r_{,q^j})\cdot \tilde{\phi}'+
\tilde{\phi}'\cdot(t\cdot ds+u\cdot r_{,q^j})\cdot \tilde{\pi}'
+\tilde{\pi}'\cdot (u\cdot s_{,q^j})\cdot \tilde{\pi}']]\; dq^j
\nonumber\\
&& [-\frac{1}{2}[\tilde{\phi}'\cdot (t\cdot r)_{,p^j} \cdot \tilde{\phi}'
+\tilde{\pi}'\cdot (u\cdot s)_{p_j}\cdot \tilde{\pi}'+
+2 \tilde{\phi}'\cdot (t\cdot s)_{,p_j}\cdot \tilde{\pi}']
\nonumber\\
&&+
[\tilde{\phi}'\cdot(t\cdot r_{,p_j})\cdot \tilde{\phi}'+
\tilde{\phi}'\cdot(t\cdot ds+u\cdot r_{,p_j})\cdot \tilde{\pi}'
+\tilde{\pi}'\cdot (u\cdot s_{,p_j})\cdot \tilde{\pi}']]
\; \times
\nonumber\\
&& d[p_j-\frac{1}{2}[\tilde{\phi}'\cdot (t\cdot r)_{,q^j} \cdot \tilde{\phi}'
+\tilde{\pi}'\cdot (u\cdot s)_{q^j}\cdot \tilde{\pi}'+
+2 \tilde{\phi}'\cdot (t\cdot s)_{,q^j}\cdot \tilde{\pi}']]
\nonumber\\
&&+[\tilde{\phi}'\cdot(t\cdot r_{,q^j})\cdot \tilde{\phi}'+
\tilde{\phi}'\cdot(t\cdot ds+u\cdot r_{,q^j})\cdot \tilde{\pi}'
+\tilde{\pi}'\cdot (u\cdot s_{,q^j})\cdot \tilde{\pi}']]
\nonumber\\
&=& \tilde{\pi}'\cdot d\tilde{\phi}'
\nonumber\\
&&+[p_j
-\frac{1}{2}[\tilde{\phi}'\cdot (t\cdot r)_{,q^j} \cdot \tilde{\phi}'
+\tilde{\pi}'\cdot (u\cdot s)_{q^j}\cdot \tilde{\pi}'+
+2 \tilde{\phi}'\cdot (t\cdot s)_{,q^j}\cdot \tilde{\pi}']
\nonumber\\
&&+
[\tilde{\phi}'\cdot(t\cdot r_{,q^j})\cdot \tilde{\phi}'+
\tilde{\phi}'\cdot(t\cdot ds+u\cdot r_{,q^j})\cdot \tilde{\pi}'
+\tilde{\pi}'\cdot (u\cdot s_{,q^j})\cdot \tilde{\pi}']]\; \times
\nonumber\\
&& d[q^j
\frac{1}{2}[\tilde{\phi}'\cdot (t\cdot r)_{,p^j} \cdot \tilde{\phi}'
+\tilde{\pi}'\cdot (u\cdot s)_{p_j}\cdot \tilde{\pi}'+
+2 \tilde{\phi}'\cdot (t\cdot s)_{,p_j}\cdot \tilde{\pi}']
\nonumber\\
&&-
[\tilde{\phi}'\cdot(t\cdot r_{,p_j})\cdot \tilde{\phi}'+
\tilde{\phi}'\cdot(t\cdot ds+u\cdot r_{,p_j})\cdot \tilde{\pi}'
+\tilde{\pi}'\cdot (u\cdot s_{,p_j})\cdot \tilde{\pi}']]
\ea
whence to second order in the perturbations we have 
(use $u\cdot r-s\cdot t=1_{L_2(T^3)}$ 
\ba \label{4.1.32}
\tilde{q}^j &=&q^j 
-\frac{1}{2}[
\tilde{\phi}'\cdot(t\cdot r_{,p_j}-r\cdot t_{,p_j})\cdot \tilde{\phi}'
+\tilde{\pi}'\cdot(u\cdot s_{,p_j}-s\cdot u_{,p_j})\cdot \tilde{\pi}'
\nonumber\\
&& +\tilde{\pi}'\cdot(u\cdot r_{,p_j}-r\cdot u_{,p_j}
+t\cdot s_{,p_j}-s\cdot t_{,p_j})\cdot \tilde{\phi}'
\nonumber\\
&=:& q^j+x^j(q,p)
\nonumber\\
\tilde{q}^j &=&p_j 
+\frac{1}{2}[
\tilde{\phi}'\cdot(t\cdot r_{,q^j}-r\cdot t_{,q^j})\cdot \tilde{\phi}'
+\tilde{\pi}'\cdot(u\cdot s_{,q^j}-s\cdot u_{,q^j})\cdot \tilde{\pi}'
\nonumber\\
&&+\tilde{\pi}'\cdot(u\cdot r_{,q^j}-r\cdot u_{,q^j}
+t\cdot s_{,q^j}-s\cdot t_{,q^j})\cdot \tilde{\phi}'
\nonumber\\
&=:& p_j+y_j(q,p)
\ea
Note that in deriving the conditions on $r,s,t,u$ one must invert 
(\ref{4.1.29}) and (\ref{4.1.32}) only up to second order 
in $\tilde{\phi}',\tilde{\pi}'$ which themselves are considered as 
of first order. Thus 
\be \label{4.1.33}
q^j=\tilde{q}^j-x^j(\tilde(q),\tilde{p}),\;
p_j=\tilde{p}_j-y_j(\tilde(q),\tilde{p})
\ee
so that up to second order 
\ba \label{4.1.34}
h_h(q,p) &=& h_h(\tilde{q},\tilde{p})
-\frac{\partial h_h}{\partial q^j}(\tilde{q},\tilde{p})\; 
x^j(\tilde{q},\tilde{p})
-\frac{\partial h_h}{\partial p_j}(\tilde{q},\tilde{p})\; 
y_j(\tilde{q},\tilde{p})
\\
&=& \tilde{h}_h-\frac{1}{2}
[
\tilde{\phi}'\cdot(t\cdot \dot{r}-r\cdot \dot{t})\cdot \tilde{\phi}'
+\tilde{\pi}'\cdot(u\cdot \dot{s}-s\cdot \dot{u})\cdot \tilde{\pi}'
+\tilde{\pi}'\cdot(u\cdot \dot{r}-r\cdot \dot{u}
+t\cdot \dot{s}-s\cdot \dot{t})\cdot \tilde{\phi}'
\nonumber
\ea
where all functions on the r.h.s. are evaluated at $\tilde{q},\tilde{p}$
and we used the abbreviation $\dot{r}=\{h_h,r\}$ etc.

Next we note that in (\ref{4.1.29}) we may replace $r(q,p)$ by 
$r(\tilde{q},\tilde{p})$ when subsituting into $h_i$ since the 
corrections would be at least of fourth order since $h_i$ is already 
of second order. Accordingly from 
(\ref{4.1.1}) with $f=a^{-3},\;\omega^2=g (-\Delta)+k, g=a,\; k=m^2 a^3$ 
in which we can also replace $a$ by $\tilde{a}$
\be \label{4.1.34a}
h_i =
\frac{1}{2}[
\tilde{\phi}'\cdot (t\cdot f \cdot t+r\cdot \omega^2\cdot r)\cdot \tilde{\phi}'
\tilde{\pi}'\cdot (u\cdot f \cdot u+s\cdot \omega^2\cdot s)\cdot \tilde{\pi}'
+2\tilde{\pi}'\cdot 
(u\cdot f \cdot t+s\cdot \omega^2\cdot r)\cdot \tilde{\phi}']
\ee
where all functions depend on $\tilde{q},\tilde{p}$. Combining
\ba \label{4.1.35}
&& h_h+h_i-\tilde{h}_h
=\frac{1}{2}
\tilde{\phi}'\cdot 
(t\cdot f \cdot t+r\cdot \omega^2\cdot r-[t\dot{r}-r\dot{t}])
\cdot \tilde{\phi}'
\nonumber\\
&& +
\tilde{\pi}'\cdot 
(u\cdot f \cdot u+s\cdot \omega^2\cdot s-[u\dot{s}-s\dot{u}])
\cdot \tilde{\pi}'
\nonumber\\
&&
+\tilde{\pi}'\cdot 
(2u\cdot f \cdot t+2s\cdot \omega^2\cdot r-
[u\dot{r}-r\dot{u}+t\dot{s}-s\dot{t}])\cdot \tilde{\pi}']
\ea
The last term is ill defined on any Fock space, hence its round bracket must 
vanish. Denote the round bracket of the second term by $l\cdot l$ which 
is supposed to be positive and is allowed to be a function of 
both $q,p$ and $-\Delta$. Then we wish 
that the round bracket of the first term takes the form 
$l(-\Delta +\tilde{m}^2) l$ where $\tilde{m}^2$ is a function of the 
homogeneous variables to be determined. This will guarantee that we can 
factor out $l^2$ from the expression of the Hamiltonian, leaving us with 
a Hamiltonian density of standard form with constant coefficients of 
$-\Delta$ so that the Hilbert-Schmidt condition is satisfied.

The simplest choice of $r,s,t,u$ is such that 1. none of them depends on
$-\Delta$ and 2. also $l$ does not depend on $-\Delta$. This 
choice is in fact unique. Namely the expression for $l$ and our assumpyion 
implies that $s=0$ and hence $r\cdot u=1_{L_2(T^3)}$.
The round bracket in the second term is then $l\cdot l=u\cdot f\cdot u$ which 
is manifestly positive. Then the condition that the round bracket 
in the last term vanishes can be solved algebraically 
\be \label{4.1.36} 
t=-l^{-1}\cdot \dot{u}\cdot l^{-1}
\ee
which leads to the final condition
\be \label{4.1.37}
l\cdot(-\Delta+\tilde{k})\cdot l=u^{-1}\cdot(l^{-1}\cdot \dot{u}\cdot{u}
\cdot l^{-1}+\omega^2+\{h_h,u\cdot t\})\cdot u^{-1}
\ee
Again using the assumption we must match the coefficients of 
$-\Delta$ on both sides and we find 
\be \label{4.1.38}
u\cdot l\cdot l\cdot u=u\cdot u\cdot f\cdot u\cdot u=g
\ee
which for the present model has the unique positive solution $u=\tilde{a}$
and thus $l^2=\tilde{a}^{-1}$. The transformed mass term becomes 
\be \label{4.1.39}
u\cdot l \cdot \tilde{k}\cdot l\cdot u
=l^{-1}\cdot \dot{u}\cdot \dot{u} \cdot l^{-1}
+k-\{h_h, u\cdot l^{-1} \cdot \{h_h,u\} \cdot l^{-1}\}
\ee
~\\
To summarise:\\
We were indeed able to make the Hamiltonian symbol well defined on the same 
Fock space for all values of the homogeneous degrees of freedom. But the 
price to pay is twofold:\\
1.\\
Due to the dependence of the mass term $\tilde{k}$ displayed in 
(\ref{4.1.39}) on both $\tilde{q},\tilde{p}$, the BOA method is no longer
available. We are {\it forced} into its generalisation, the SAPT scheme.\\
2.\\
The mass term is not manifestly positive, it is generically 
{\it indefinite} and there is no freedom left 
to change this without making the coefficients $r,s,t,u$ also 
depend on $-\Delta$. Whether this can be 
improved by exploiting the complete freedom in thos 
coefficients will be left for future research.\\
3.\\
In this respect, we draw the attention of the reader to reference 
\cite{25}. There the starting point is indeed a Hamiltonian of 
second order in the inhomogeneous degrees of freedom with 
standard form up to a prefactor depending on the homogeneous degrees
of freedom, except that the mass squared is a generic function of 
the homogeneous degrees of freedom. The Mukhanov-Sasaki Hamiltonian is a
prominent example. Hence we are precisely in the situation 
arrived at above after our canonical transformation (exact up to second 
order). The analysis of \cite{25} investigates the most general Fock 
representation, labelled by the homogeneous variables, that  
supports such a Hamiltonian and at the same time provides a canonical 
transformation of the homogeneous sector to variables which directly
commute with the associated {\it annihilation and creation variables}.
This has the adavantage that the Hilbert-Schmidt condition is 
trivially solved because the annihilation and creation operators 
do not depend on the transformed homogeneous degrees of freedom.
As such, the strategy is similar in spirit to the present one although 
the details are different.    
One finds that in this case an algebraic solution is no longer possible,
rather one must solve a system consisting of two non-linear (but semi-linear) 
first order PDE's for {\it complex coefficient operators} coming from the 
Hamiltonian vector field of $h_h$ to guarantee that all conditions are met, 
including
the positivity of the mass term. One of the conditions is equivalent 
to the fixed point equation of the adiabatic vacua construction 
\cite{7}, the other determines an otherwise free phase. While those   
PDE's are well posed and can be solved in principle by the method of 
caracteristics, it is generically very hard to to solve the system explicity
given the detailed form of $h_h$
which, however, is a prerequisite to quantise the homogeneous sector as 
well. Thus for the purpose of the papers in this series, we stick to the 
method sketched above, although the possibility to guarantee the positivity
of the mass squared term using the methods of \cite{25} is very attractive 
in view of the complications that arise for negative mass squared terms 
discussed in the next subsection. 

There is is also another, independent
reason for why the approach of \cite{25} is especially attractive: Since 
annihilation and creation operators commute with the operators of the 
homogeneous sector, the latter operators preserve the domain of the 
inhomogeneous part of the Hamiltonian. This is not necessarily the case when 
just guaranteeing the Hilbert-Schmidt condition. To see 
this suppose that the symbol $\kappa'$ in (\ref{4.1.19}) is of Hilbert-Schmidt
type and only depends on $a$, then 
the vector $H_i P\Omega,\; H_i=W(h_i)$ is given by (recall (\ref{4.1.21})
\be \label{4.1.40}  
H_i P\Omega=-i\hbar\int_\sigma d^3x B(x)^\ast (\hat{\omega}\cdot 
K'\cdot B)^\ast(x)
\Omega
\ee
but the symbol of $\hat{\omega}\cdot K'$ is up to a factor given by 
$\omega(a) \frac{d}{da} \ln(\omega(a))=d\omega(a)/da$ 
which decays only as $k_I$ even if the coefficient of $-\Delta$ in 
$\omega(a)^2$ is independent of $a$. By itself this is not a problem
because we want to consider the spectrum of $H=W(h)$ rather than $W(h_i)$
which does not require to have the commutator $[P,H_i]$ defined on the Fock
space, however, it would be a convenient property to have (recall
that once $H$ can be constructed as a self-adjoint operator the existence
of a dense and invariant domain is granted \cite{25a})
Again, 
to have a domain left invariant by the operators of the 
homogeneous sector could possibly granted within 
the context of this paper if we considered general $r,s,t,u$ 
and in particular make the derivative of the function 
$l\cdot (-\Delta +M^2)\cdot l$
with respect to the homogeneous variables decay at least as $k_I$.

In view of this discussion we hope to come back to the formalism of 
\cite{25} adapted to the present context in a future publication.

\subsection{Indefinite mass squared operators}
\label{s4.3}

The discussion of the previous section applies to rather generic
second order Hamiltonians. More generally one may have several matter 
or geometry species 
e.g. scalar, vector, tensor and spinor modes \cite{26}.
We can apply for each inhomogeneous species an individual 
canonical transformation parametrised by $r_s,s_s,t_s,u_s$ 
where $s$ labels the species and transform the 
symplectic potential  
for each of the species simulatenously the effect of which is that 
simply the $x^j_s, y_j^s$ corrections for all species add up. Since we 
perturb $h_h$ only linearly in $x^j, y_j$ and since 
in each second order Hamiltonian we can drop the $x^j,y_j$ corrections,
the species contributions never
mix up to second order perturbation theory. 

Accordingly we can consider the hamiltonian symbols to be well defined 
on the corresponding Fock spaces and the Hilbert-Schmidt conditions are all
solved. However, as concluded in the previous subsection, the mass squared 
terms $M^2_s$ for each of the species are generically not positive on the 
entire phase space. In what follows we present several strategies for 
how to deal with this problem, none of which is entirely satisfactory
as they either lead to instabilities or contain ad hoc elements. Since 
this is a new situation, the discussion will be mostly exploratory.
\begin{itemize}
\item[1.]
One could exploit the full freedom in those transformations beyond the 
restricted 
Ansatz of the previous subsection to try to make positivity manifest.
This possibility will be explored elsewhere. In that respect it should 
be mentiooned that there seems to be substantial freedom in the choice of
the functions $u,r,s,t$ and thus the regime of the phase space
where the mass squared functions become negative depends on that 
freedom. In that sense, that region should not be of any physical 
significance and it would be most natural to get rid of it, thus 
restricting the freedom in the choice of the canonical transformations 
by a physically motivated criterion.\\
\item[2.]
One could restrict the classical phase space of the homogeneous degrees of 
freedom 
to the set of points $(q,p)$ where $M^2_s(q,p)\ge 0$ for all $s$. This 
restriction can be achieved by definining new variables $v_s$ and 
to set $v_s^2=M^2_s$. An especially nice situation is when the 
the $M^2_s$ have mutually vanishing Poisson brackets between 
them as we then can consider 
them as action variables and determine the corresponding angle 
variables as conjugate ones. This is in particular possible for a single 
field species as in (\ref{4.1.1}) but already fails when we have 
tensorial and scalar field modes present at the same 
time. More generallly, we may
be able to write $M_s^2$ in the form 
\be \label{4.3.1}
M_s^2(q,p)=F_s^2(q,p) N_s^2(q,p)
\ee
where now $F_s(q,p)$ is a real function and $N^2_s$ is still indefinite 
but the $N_s^2$ are mutually commuting for all $s$ for which 
$N^2_s$ is indefinite. Then we may apply the action angle prescription
to $v_s^2=N_s^2$ (assuming that the number of homogneous pairs is at least as 
large as the number of indefinite mass squared terms). In the most 
general case we solve   
the equations $v_s^2=M_s^2(q,p)$ for some homogeneous momenta 
$p_s=F_s(v,c;z)$ (this may involve choosing branches)
where $c$ is the collection of the $q^s$ conjugate to
the chosen $p_s$ and $z$ stands for the remaining canonical pairs. The 
variables $v,c,;z$ then coordinatise a new phase space with induced 
symplectic structure and while they are 
not canonical coordinates for it, they are supposed to 
have full range in some $\mathbb{R}^{2m}$ in contrast to 
the $p_s$. We then must pass to suitable Darboux coordinates and 
hope that they are global in order that we may apply Weyl quantisation.

To illustrate this, we compute the mass squared opertor for the 
model (\ref{4.1.1}) for which the homogeneous piece of the Hamiltonian 
reads ($c$ is a positive constant)
\be \label{4.3.2}
h_h=-c \frac{p^2}{a}+\Lambda a^3+\frac{1}{2}(\frac{\pi^2_0}{a^3}+
m^2 a^3 \phi_0^2)
\ee
It follows 
\be \label{4.3.3}
\dot{a}=\{h_h,a\}=-2c \frac{p}{a},\;
\dot{p}=\{h_h,p\}=-c \frac{p^2}{a^2}-3\Lambda a^2+\frac{3}{2}
(\frac{\pi_0^2}{a^4}-m^2 a^2 \phi_0^2)
\ee
Thus from (\ref{4.1.39}) with $u=a,f=a^{-3},l\cdot l=a^{-1}$
\be \label{4.3.4}
a \tilde{k}=\frac{\dot{v}^2}{9v}+m^2 v-\frac{\ddot{v}{3}},\;v=a^3
\ee
we find 
\be \label{4.3.5}
\tilde{k}=[m^2-6c \Lambda-3c m^2 \phi_0^2] a^2+3c\frac{\pi_0^2}{a^4}
-2c^2\frac{p^2}{a^2}
\ee
which is clearly indefinite. To illustrate more clearly the procedure, 
suppose that we would have treated the homogeneous mode of the scalar
field on equal footing with the inhomogeneous ones so that the 
$\phi_0,\pi_0$ dependent terms are missing from (\ref{4.3.2}) and thus 
(\ref{4.3.5}) which however is still indefinite even for 
tiny cosmological constant so that $M^2:=m^2-6c \Lambda>0$.   
We define a new canonical pair $(b,q)\in \mathbb{R}^2$ and functions 
$a,p$ of $b,q$ by (set $\delta^2:=2 c^2/M^2$)  
\be \label{4.3.6}
a^2:=b^2+\delta^2 \frac{q^2}{b^2},\; p:=\frac{a}{b} q
\ee
Note that in cosmology $a>0$ so that the square root in (\ref{4.3.6}) 
has only one branch and thus (\ref{4.3.6}) is uniquely defined away from 
$b=0$. To see that this is locally a canonical transformation we 
compute 
\be \label{4.3.7}
\{p,a\}=\{\frac{a}{b} q,a\}
=a\{\frac{q}{b},a\}
=\frac{1}{2}\{\frac{q}{b},a^2\}
=\frac{1}{2}\{\frac{q}{b},b^2+\delta^2 (\frac{q}{b})^2\}
=\frac{1}{2b}\{q,b^2\}=1
\ee
and thus 
\be \label{4.3.8}
\tilde{k}=M^2 b^2
\ee
is manifestly positive. However, (\ref{4.3.6}) restricts the original range 
$(a,p)\in \mathbb{R}_+\times \mathbb{R}$ to the set of pairs $(a,p)$
with $a^4\ge \delta^2 p^2,\;p\in \mathbb{R}$. 

In general, even if possible, the canonical transformations involved 
worsen the degree of non-polynomiality of the symbol $h(z)$
with respect to the homogeneous degrees of freedom $z=(q,p)$.  
\item[3.] The third possibility to is to take the indefinite mass terms
$\tilde{k}$ 
seriously as the stand. Accordingly, for certain ranges of the 
homogeneous variables $(q,p)$ the inhomogeneous symbol $h_i$ defines a
{\it quantum field theory of tachyons}! A possibility to deal with 
the tachyonic instabilty was suggested in \cite{17}: We construct the 
Fock space ${\cal H}_{(q,p)}$ 
as before but we only allow those modes corresponding to eigenfunctions 
$f_I,\;I\mathbb{Z}^3-\{0\}$ of $-\Delta$ witheigenvalue 
$k_I^2+\tilde{k}(q,p)\ge 0$. 
Accordingly, the more negative $\tilde{k}$ becomes the larger the 
infrared cut-off on the admisaable modes. Specifically, for the 
example (\ref{4.3.5}) with $\phi_0=\pi_0$ we find that  $\tilde{k}$
gets very negative for $p^2/a^2\to \infty$. If we interprete this as
$\propto \dot{a}^2$ and take a baryon or radiation dominated universe then 
this certainly diverges at the classical big bang. 

For the SAPT theory this has the following consequence:\\
If we take a torus of length $L$ in all directions then 
$k_{I}^2=k_L^2 ||I||^2,\; k_L=2\pi/L$. Let $z=(q,p)$ and 
$\tilde{k}(z)$ be the indefinite mass squared term. Let 
$S_\pm$ be the subsets of the slow phase space defined by $\tilde{k}(z)\ge 0$ 
and $\tilde{k}(z)<0$ respectively. We can enumerate the spectrum
$E_n(z)$  
of the Hamiltonian symbol $h_(z)$ by a mode number $M$, positive 
intergers $N_1,..,N_M$ and mutually distinct positive numbers 
$r_1<..,<r_M$ with $r_k=||I_k||,\; I_k\in \mathbb{Z}^3-\{0\}$.
The spectral value is given by 
\be \label{4.3.9}
E_n(z)=\sum_{k=1}^M\; N_k\; \sqrt{k_L^2 r_k^2+\tilde{k}(z)} 
\ee
The mode number configurations which give rise to the same $E_n(z)$ 
determine the degeneracy of $E_n(z)$. First we see that varying 
the $M,N_k,r_k$ does not leave (\ref{4.3.9}) invariant (a.e. wrt
$z$) as otherwise the 
numbers $\sqrt{k_L^2 r^2+\tilde{k}(z)}$ would be linearly dependent 
over the positive rationals which is not the case a.e. It follows 
that the only degeneracy lies in choosing the $I_k$ with given $r_k$
for which there are 8 possibilities, thus the degenracy of $E_n(z)$
is $8^M$ with $n=n(M,\{N_k\},\{r_k\}$ independent of $z$ when 
$z\in S_+$.  

However, for $z\in S_-$ we compute $r(z)^2:=-\tilde{k}(z)$ and 
can only allow the energy band $n$ with $r_1\ge r(z)$. It follows that 
the eigenvalue $E_n(z)$ simply does not exist when $r_1<r(z)$. 
Consequently, also the eigenstates $e_n(z)$ and the standard vectors $b_n$
that enter the Moyal projectors and unitarities $\pi_{n,0}(z),u_n(z)$ 
are simply deleted, the Fock space ${\cal H}_f(z)$ is the subspace 
of ${\cal H}_f$ spanned by the $e_n(z)$ with $r_1\ge r(z)$. 
In other words, the function $z\mapsto E_n(z)$ for 
given $n$ has a discontinuity at the surface $r_1=r(z)$ in the phase space.
This can be problematic when computing the the Moyal products which ask
to take derivatives with respect to $z$, a possibility being to take 
the one sided derivatives only. 
\item[4.] In an ad hoc manner, one could restrict the phase space integral 
that enters the Weyl quantisation to $S_+$, i.e. one multiplies 
all symbols such as $h(z)$ with $\chi_{S_+}(z)$ where $\chi_{S_+}(z)$ 
denotes the characteristic function of $S_+$. This is again not 
differentiable 
and thus it would be more appropriate to substitute $\chi_{S_+}$ by a 
mollified version of it (i.e. a smooth function that is zero in 
$S_-$ and smoothly reaches unity within $S_+$ 
in an arbitrarily small neighbourhood 
of the boundary $\partial S_+$) 
in order that the Moyal product is meaningful.
Of course, the quantum theory then will depend on that mollification
which introduces ambiguities and technical challenges as the mollifier
is a highly non polynomial function of $z$.
\item[5.] We could consider a mode decomposition of $h_i(z)$ and 
for $z\in S_-$ write $h_i(z)=h^+_i(z)+h^-_i(z)$ where $h^+_i$ is the 
contribution 
from all modes $I$ with $k_I^2\ge r(z)$. Then $h^+_i(z)$ is quantised as 
before and $h^-_i(z)$ is a finite sum of flipped harmonic oscillators
of the type $p^2-\omega^2 q^2, \omega^2>0$. The difference with item 
[3.] is that 
we do not discard $h^-_i$. The spectrum of a flipped harmonic 
oscillator is purely of the absolutely continuous type \cite{15} and 
thus the spectrum of $h_i(z)$ is drastrically changed when we transit
from $S_+$ to $S_-$ with corresponding consequences for the SAPT scheme.
Besides, such a theory would be unstable.
\end{itemize}
For the model (\ref{4.1.1}) strategy [2.] seems to be most promising
as we will see in another paper of this series.

\subsection{Non-polynomial operators}
\label{s4.4}

The purely homogeneous piece $h_h$ of the Hamiltonian is non-polynomial
in the scale factor $a$ and contains inverse powers of it. The mass squared 
corrections coming from the canonical transformation performed in section 
(\ref{s4.2}) contains derivatives of $h_h$ and increases that negative 
power. Furthermore, the adiabatic corrections contain additional 
derivatives of $h_h$ of aribtrary order coming from the 
Moyal product and thus introduces furteher arbitrarily 
negative powers of $a$. Even worse, after the mass squared corrections 
we potentially also find inverse powers of arbitrarily high order
in the momentum $p$ conjugate to $a$, the Mukahnov Sasaki mass term being a 
prominent example.

It transpires that it would be desirable to have at one's disposal 
a dense set of vectors which is invariant under any of the operators 
corresponding to $a^n,p^n, \; n\in \mathbb{Z}$. In LQC one deals 
with negative powers of $a$ by using a representation inspired by 
the representation used in the full LQG theory such that the spectrum
of $a$ is pure point rather than absolutely continuous and thus 
a commutator between fractional powers of $a$ and Weyl elements for $p$ 
is both densely defined and introduces the desired negative powers $a$.
This comes at the price that the operator corresponding to $p$ does not 
exist and one thus needs to approximate it by polynomials in 
Weyl elements. However, negative powers of $p$ would then also need 
to be approximated by inverse polynomials of Weyl elememts and these 
are not in the domain of $a$ so that for our purpose the representation
chosen in LQC is of no direct advantage.

We thus advocate to take an unbiased point of view and ask whether
it is possible to choose the above desired domain directly in the 
Schr\"odinger represenation, the advantage being that the operators 
corresponding to $a,p$ exist. We found the following answers:
\begin{Theorem} \label{th1}~\\
Let ${\cal H}=L_2(\mathbb{R},x)$ be the Schr\"odinger representation 
of $q,p$ as operators $(Q\psi)(x)=x\psi(x),\;(P\psi)(x)=i d\psi(x)/dx$.
i.\\
There exists a dense and invariant domain $D$ for the operators $Q^n\;
P^m,\;
n\in \mathbb{Z},\; m\in \mathbb{N}_0$ consisting of smooth functions of 
rapid decrease both at $x=0$ and at $x=\pm \infty$.\\
ii.\\
$D$ is spanned by functions $b_n,\; n\in \mathbb{Z}$ whose inner products 
can be computed analytically in closed form. Correspondingly an orthonormal
basis can be be constructed by the Gram-Schmidt procedure.\\
iii.\\  
Let $F$ be a function such that $F^{-1}$ is 
a polynomially bounded function both 
in terms of $x$ and $x^{-1}$ and smooth except possibly at $x=0,\pm\infty$.
Let $f_1,..,f_N$ be polynomials in $x$.    
Then there exists a common domain $D_L(F)\subset D$ for the operators of 
item i. and 
of the operators corresponding to the symols $|F(q)|^2\;
f_k(q) p^{-k},\; k=1,..,N$ 
in suitable symmetric orderings where $L$ depends on both $N$ and the degree 
of the polynomials $f_k$. 
\end{Theorem}
The proof of this theorem can be found in \cite{24}. Note that 
$P^{-1}$ is a symmetric operator with distributional kernel 
\be \label{4.4.1}
(P^{-1} \psi)(x)=-\frac{i}{2\hbar}\int_{\mathbb{R}}\; dy \; {\rm sgn}(x-y)
\psi(y)
\ee
The domain of $P^{-1}$ must be carefully chosen: Even if $\psi$ is a 
Schwartz function, while $P^{-1}\psi$ is smooth, it may not be of 
rapid decrease any more at infinity. Likewise, it is a simple collary
that a dense and invariant domain
for $P^n Q^m,\; n\in \mathbb{Z},\; m\in \mathbb{N}_0$ is given by 
the Fourier transform of the functions of item i. but that Fourier 
transform is not necessarily of rapid decrease in $x$ any more. This is
why the statement of item iii. is significantly weaker, in particular, 
$D_L(F)$ is not an invariant domain for the list of operators stated 
and it is presently not clear whether it is dense.  
It is however certain that there exists no function in $D$ 
orthogonal to $D_L(F)$. 

The idea for defining the rather singular symbols that we encounter 
in the homogeneous sector of quantum cosmology is thus as follows
(provided that we can factor out a suitable $|F|^2$ as described above):
At any order of the adiabatic expansion the terms that involve negative 
powers of $p$ are of the form 
described in item iii.  and are finite in number. Thus we use the ordering
alluded to in item iii. and the domain described there. The other terms 
not involving negative powers of $p$ are also defined on that domain 
since $D_L(F)\subset D$.

\section{Conclusion and Outlook}
\label{s5}

In the present first paper of this series we provided the tools with which 
we will intend to improve on the treatment of backreations in 
quantum cosmology. Thus we prepared the ground to 
approach the various models that are being treated in the subsequent 
papers of the series. \\
\\
The plan of the subsequent papers in this series is as follows:\\
\\
In the second paper \cite{28} we treat the two quantum mechanical models 
labelled as models I and II in section \ref{s4.2}. Model I is a standard 
QM problem consisting of polynomially coupled slow anharmonic and fast harmonic 
osccillator which mimicks the situation of model II and serves 
to illustrate the formalism. Model II  
considers the purely homogeneous cosmological sector, i.e. 
homogeneous geometry coupled to a homogeneous inflaton. 
In suitable variables this model can be dispalyed as an inverted slow 
harmonic osciallator non-polynomially coupled to a fast standard harmonic 
oscillator. The adiabatic 
parameter squared is here the ratio of coupling constants for gravity and 
the inflaton which we assume to be very small. The adiabatic parameter 
can also be written as the ratio of corresponding 
inflaton and Planck mass scales and thus is very tiny if we consider the 
latter to be of the order of the mass scales that appear in the current 
standard 
model of elementary particle physics. That parameter will also organise 
the adiabatic perturbation expansion of the third and fourth paper.

In the third paper \cite{29} 
we consider as matter content an inflaton and Gaussian 
dust. The usual Hamiltonian constraint is now a physical Hamiltonian as 
shown in \cite{5} and the full constraints, not only their perturbations are 
already solved, all metric and inflaton degrees of freedom 
are physical observables. We expand the physical Hamiltonian to second order
in the inhomogeneous modes leading to three scalar, one vector and two tensor 
modes. For simplicity we consider only quantisation of the inflaton field,
i.e. we drop all metric perturbations and keep only the homogeneous metric 
degrees of freedom.

Finally in the fourth paper \cite{30} 
we consider as matter content just the inflaton 
field and follow closely \cite{8} in order to extract the gauge invariant
observables of which there is the Mukhanov-Sasaki field and the tensor 
mode (primordial gravitational wave). 

In all papers we compute the backreaction effects to second order 
in the adiabatic parameter thus displaying their existence and potential 
phenomenological importance that we will explore in a forthcoming 
publication. Note that the model of the second paper can be considered as 
the purely homogeneous truncation of both the model of the 
third and the fourth paper respectively, 
just that in the first case it is to be consideered as a dynamical system 
with true Hamiltonian, in the second case that Hamiltonian is constrained 
to vanish. Accordingly, for the second paper we are interested in the 
full spectrum of the Hamiltonian which in appropriate variables can be 
considered as a harmonic oscillator non-polynomially coupled to an 
inverted harmonic oscillator \cite{15}.\\   
\\
\\
\\
{\bf Acknowlegdements}\\
\\
We would like to thank Beatriz Elizaga Navascues for drawing our 
attention to the 
hybrid LQC strategy to use canonical transformations in second order 
cosmological perturbation theory in order to satisfy the Hilbert-Schmidt 
condition in the corresponding quantum field theory. S.S. thanks the
Heinrich-B\"oll Stiftung for financial and intellectual support and the
German National Merit Foundation for intellectual support.

}

\end{document}